\documentclass[aps,twocolumn,prx,amsfonts,showpacs,superscriptaddress,longbibliography]{revtex4-2}
\usepackage{graphicx}
\usepackage{bm}
\usepackage{hyperref}
\usepackage{amsmath}
\usepackage{amssymb}
\usepackage[table]{xcolor}
\usepackage{array}
\usepackage{multirow}
\usepackage[caption=false]{subfig}
\usepackage{physics}
\usepackage{booktabs}
\usepackage{soul}
\usepackage{amsthm}
\usepackage{bbm}
\usepackage{multirow}
\usepackage[scr=boondoxo,frak=boondox,scrscaled=1.05]{mathalfa}
\usepackage{amsmath}
\usepackage{xspace}

\def\ket#1{|#1\rangle}
\def\braket#1#2{\langle#1|#2\rangle}

\def\r{{\boldsymbol{r}}}
\def\x{{\boldsymbol{x}}}
\def\k{{\boldsymbol{k}}}
\def\p{{\boldsymbol{p}}}
\def\q{{\boldsymbol{q}}}

\def\G{{\boldsymbol{G}}}
\def\Q{{\boldsymbol{Q}}}

\def\a{{\boldsymbol{a}}}

\def\L{{\boldsymbol{L}}}

\newcommand{\bQ}{\bm {Q}}

\newcommand{\rlg}{R$L$G\xspace}

\usepackage{ulem}
\begin{document}

\title{Quantum Oscillations across Hole Crystallization in Rhombohedral Graphene}

\author{Daniele Guerci}

\affiliation{Department of Physics, William \& Mary, Williamsburg, VA-23187, USA}

\author{Enrico Rossi}

\affiliation{Department of Physics, William \& Mary, Williamsburg, VA-23187, USA}

\begin{abstract}

Recent quantum-oscillation measurements in rhombohedral graphene reveal an abrupt increase in the dominant frequency and the emergence of additional frequencies as the carrier density is lowered toward superconductivity. Noninteracting band models can account for multiple frequencies but do not explain the upward jump. We show that these features can arise from two competing fates of the inner hole pocket in an annular Fermi sea: collapse into a simply connected Fermi liquid or crystallization into an anticrystal, in which localized holes coexist with itinerant electrons. Pocket collapse contracts the outer Fermi contour at fixed density, whereas hole crystallization preserves a larger contour that supports magnetic-breakdown orbits. This contrast provides a possible explanation for the frequency jump, while additional reconstructed orbits produce a multitone spectrum. Our results support the anticrystal as a candidate normal state near superconductivity in rhombohedral graphene.

\end{abstract}
\maketitle
\date{\today}

\textit{Introduction ---} Wigner crystals~\cite{Wigner1934} and Fermi liquids~\cite{Landau1956,Landau1957Oscillations} highlight the particle-like and wave-like character of interacting electrons in two dimensions~\cite{Ceperley1980,Tanatar1989,Attaccalite2002,Drummond2009}: electrons localize on a lattice in the crystal, whereas delocalized quasiparticles form a Fermi surface in the liquid.
Between these limits, theoretical proposals suggest phases in which crystalline order and metallic behavior coexist ~\cite{Brazovskii1975,Spivak2004,Falakshahi_2005,kim2023dynamicaldefectstwodimensionalwigner,Kim2024PRB}.

Rhombohedral $L$-layer graphene (R$L$G) has emerged as an almost ideal platform for exploring this coexistence, thanks to its low disorder and gate-tunable carrier density ($n$) and band dispersion. 
In the spin- and valley-polarized quarter-metal (QM) regime, transport experiments reveal metallic states with signatures of electronic crystallization~\cite{han2026evidencemetallicwignercrystal,zhou2026competingordersdrivenwigner}. More recently, direct imaging of metallic crystals in the valence band has provided further evidence for this coexistence~\cite{sun2026realspacevisualizationemergentelectron}.
Related states emerge in Hartree--Fock (HF) calculations~\cite{Dong_2026,Feng2026,deng2026metallicelectroncrystalsrhombohedral} and persist when correlations beyond HF are included~\cite{abouelkomsan2026rhombohedralgraphenetalecrystals,kim2026shapewignercrystalshole}. 
The observation of superconductivity in the same QM regime~\cite{Han2025,dutta2026reconfigurablechiralsuperconductivity,nguyen2026coexistingchargedensitywave,qin2026stripeordermetallicsuperconducting,hua2026multiknobswitchablechiralsuperconductivity,nguyen2025hierarchysuperconductivitytopologicalcharge,zheng2026chiralsuperconductorscompetingstates} raises the possibility that these metals with crystalline charge order provide the normal state from which superconductivity emerges. 
Establishing such a connection requires understanding how charge order reshapes the Fermi surface and identifying its experimental signatures.

Shubnikov-de-Haas (SdH) quantum oscillations of the longitudinal resistance ($R_{xx}$) probe the Fermi surface, providing key information about the superconductor's normal state.  
Recent measurements in the QM regime~\cite{kalantre2026} reveal, after careful background subtraction, an abrupt change in the oscillation spectrum near the onset of superconductivity: as the carrier density is reduced through a critical value $n_c$, the dominant frequency jumps upward and additional frequencies appear. 
The upward jump signals larger quasiparticle orbits despite the decreasing carrier density. 
While the multiple frequencies can be understood within a non-interacting description~\cite{zhao2026magneticbreakdownanomalousquantum}, the discontinuous increase in the dominant frequency remains unexplained within that framework.

In this Letter, we show that interactions in annular Fermi seas give rise to this phenomenology through a transition from a Fermi liquid to an \textit{anticrystal}, a metallic state in which a triangular crystal of localized holes coexists with itinerant electrons~\cite{abouelkomsan2026rhombohedralgraphenetalecrystals}.
%
%
In the liquid, interactions eliminate the inner hole pocket, leaving a simply connected Fermi surface.
In the anticrystal, the periodic potential of the hole crystal reconstructs the electron-like Fermi surface into disconnected segments. 
Magnetic breakdown connects these segments into a large orbit whose area approaches that enclosed by the bare outer Fermi contour and exceeds the Fermi-surface area in the liquid. 
This difference provides a mechanism for the upward jump in the dominant SdH frequency across the transition. 
Additional semiclassical orbits produce a multitone spectrum. 
Using HF calculations and a semiclassical analysis, we establish this mechanism and identify the anticrystal as a candidate normal state for superconductivity in the QM regime of R$L$G.

\textit{Model ---} We consider spin- and valley polarized electrons described by the Hamiltonian: 
\begin{equation}\label{hamiltonian_ground_state}
    \hat H=\sum_{\k}\epsilon_c(\k)\hat c^\dagger_{\k} \hat c_{\k}+\frac{1}{2A}\sum_{\q}V(\q):\hat\rho(-\q)\hat \rho(\q):,
\end{equation}
where $\epsilon_c(\k)$ is the dispersion relation of the conduction band above charge neutrality of R$L$G, $\k$ is unconstrained, $\hat c_{\k}$ ($\hat c^\dagger_{\k}$) is the annihilation (creation) operator of electrons with momentum $\k$ and wave function $\ket{u_{\k c}}$.
In our analysis, we employ $V(\q)=2\pi \ell E_{C}/|\q|$ where $E_C=e^2/(4\pi \epsilon\epsilon_0 \ell)$ is the Coulomb interaction at a relative distance $\ell=\hbar v_F/t_1\simeq 1.738$nm with $v_F$ Dirac velocity and $t_1$ interlayer-intersublattice hopping, and relative dielectric constant $\epsilon$. We use model parameters from Refs.~\cite{Yu2025,Jonah2024,Kwan2025}, details are provided in the Supplementary Material (SM)~\cite{supplementary}. 
Finally, the projected density operator $\hat\rho(\q)=\sum_{\k}\Lambda_{\k,\k+\q}\hat c^\dagger_{\k}\hat c_{\k+\q}$, $\Lambda_{\k,\k+\q}=\braket{u_{\k c}}{u_{\k+\q c}}$.

\begin{figure}
    \centering
    \includegraphics[width=\linewidth]{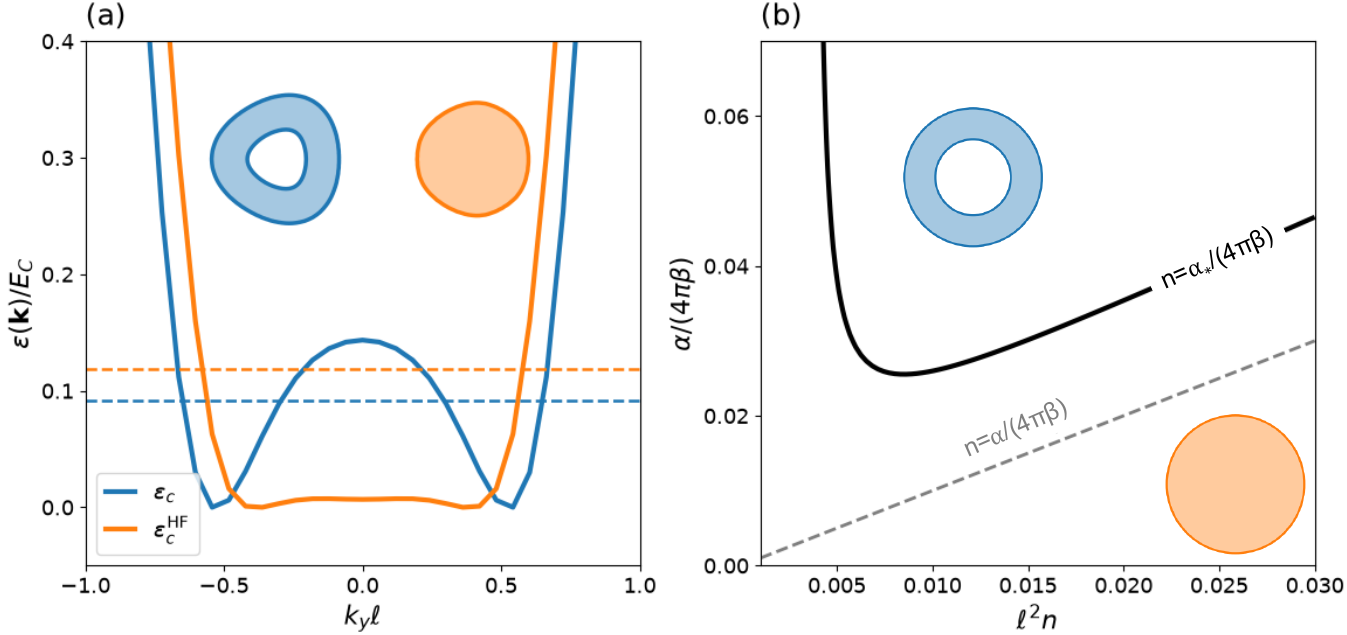}
    \caption{\textbf{Fermi surface renormalization in annular Fermi sea:} 
    (a) Interaction-induced band renormalization in the quarter-metal Fermi liquid for $u_D=48\,\mathrm{meV}$, $n=0.88\times 10^{12}\,\mathrm{cm}^{-2}$, $L=4$, and $\epsilon=15$. The bare dispersion and Fermi surface (blue) are compared with their interacting counterparts (orange). (b) Local stability of the rotationally symmetric liquid in the ``Mexican-hat" model for $e^2\ell/(4\pi\epsilon\epsilon_0\alpha)=0.6$. The solid line marks the instability of the simply connected Fermi sea toward opening an inner hole pocket. }
    \label{fig:band_renormalization}
\end{figure}

Our model describes Coulomb-interacting electrons in a continuum but differs from a conventional 2DEG in key respects:
(i) spin- and valley-polarization break time-reversal symmetry and give rise to orbital ferromagnetism; (ii)  the geometry of the underlying wave functions enters the Coulomb interaction through the complex overlap form factors $\Lambda_{\k,\p}$; (iii) while the interaction strength in a conventional 2DEG is characterized by $r_s=1/(a_B^*\sqrt{\pi n})$ with $a^*_B=4\epsilon\epsilon_0\pi\hbar^2/(m_*e^2)$, the displacement field $(u_D)$  provides an additional tuning parameter by reshaping the conduction-band dispersion $\epsilon_c(\k)$, thereby changing the relative strength of Coulomb interactions and kinetic energy even at fixed density~\cite{abouelkomsan2026rhombohedralgraphenetalecrystals}.

At sufficiently large displacement fields, the non-interacting dispersion develops a ``Mexican-hat'' profile, supporting an annular Fermi sea over a range of carrier densities. This sea is bounded by an outer electron-like and an inner hole-like Fermi surface, as shown in Fig.~\ref{fig:band_renormalization}(a).

\begin{figure}
    \centering
    \includegraphics[width=\linewidth]{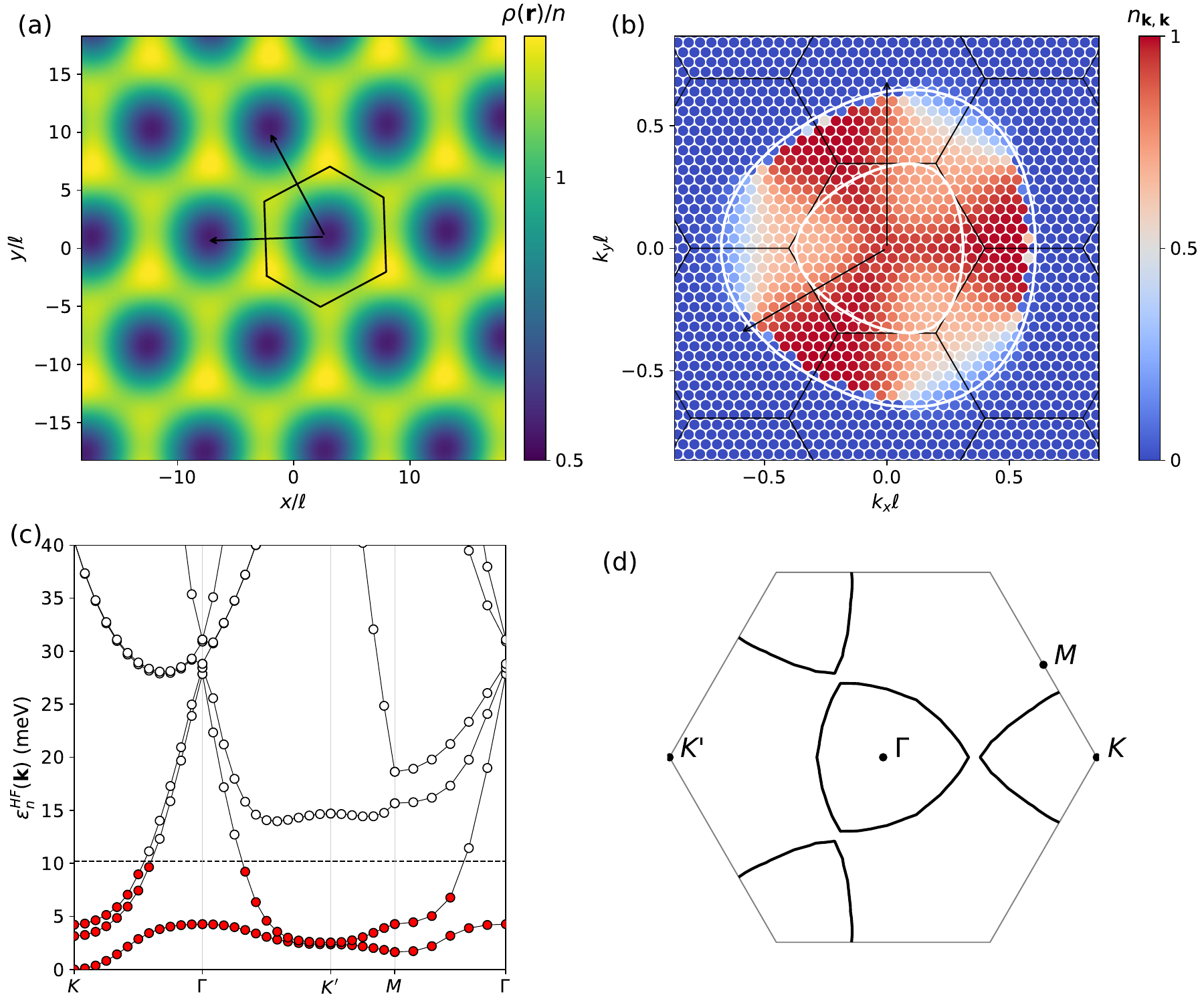}
    \caption{\textbf{Charge density, occupation, bands and Fermi surface of the anticrystal:} (a) Charge density $\rho(\r)$ of the anticrystal featuring $\nu\approx2.2$ electrons per unit cell. (b) Quasiparticle occupation $n_{\k,\k}$ where solid white lines show the noninteracting Fermi surface. Black lines outline the real-space unit cell in (a) and the Brillouin zone in (b). (c,d) Band structure and Fermi surface, respectively, in the Brillouin zone of the hole crystal.
    Calculations are performed for $N=516$ electrons (unrestricted) and $N=900$ (restricted) at $u_D=48$meV, $n=0.8\times10^{12}$cm$^{-2}$, $L=4$ and $\epsilon=15$.}
    \label{fig:AC}
\end{figure}

\textit{Fermi surface reshaping and hole crystallization in annular Fermi seas ---}  
An annular Fermi sea supports several competing many-body phases in the low-density limit, including anisotropic~\cite{Berg_2012,ge2026nematic} and unconventional Wigner crystals~\cite{joy2023wigner,joy2025chiral,ge2026nematic,kim2026exchange,kim2026shapewignercrystalshole,MoralesDuran2026}, as well as nematic Fermi liquids~\cite{Jung_2015,Chunli2023,parramartínez2025nematicpartiallypolarizedphases}. 
Nontrivial band topology further enriches this landscape by enabling topological crystalline phases~\cite{Dong_2024,dong2024theory,zhou2024fractional,zeng2024,Tan2024,dong2024stability,Kwan2025,Guo2024,Tomo2025,soejima2024anomalous,Sheng2024,Valenti_2025,Zverevich2026,bernevig2025berry}. 
Here, we focus on electron densities $n=0.6$–$0.9\times10^{12}\mathrm{cm}^{-2}$ relevant to R$L$G, where electrons near the outer Fermi surface remain itinerant, while interactions can reshape or destabilize the inner hole pocket.

To examine this competition, we first consider how interactions reshape the annular Fermi sea within the liquid phase.
At fixed carrier density, Luttinger's theorem constrains only the area between the inner and outer Fermi surfaces, but allows their individual radii to vary. The HF energy of the liquid per particle is
\begin{equation}\label{hartree-fock_liquid}
    E/N =\frac{1}{n}\int_{\k}\epsilon_c(\k)n_\k
    -\frac{e^2}{4 \epsilon\epsilon_0  n}\int_{\k}\int_{\p}
    \frac{n_{\k}n_{\p}}{|\k-\p|}|\Lambda_{\k,\p}|^2,
\end{equation}
where $n_{\k}=\langle\hat c^\dagger_{\k}\hat c_{\k}\rangle$ is the occupation number obtained by minimizing $E$ and $\int_{\k}=\int d^2\k/(2\pi)^2$. 
Fig.~\ref{fig:band_renormalization}(a) shows how interactions reshape an annular Fermi sea into a simply connected one.
This reshaping is driven by Coulomb exchange, which, at sufficiently high densities in the liquid regime, acts as an effective surface tension in momentum space, shrinking and ultimately eliminating the inner hole pocket.

To understand this mechanism, we consider a simplified ``Mexican-hat'' dispersion~\cite{geier2024chiraltopologicalsuperconductivityisospin,Kim_2026,kim2026shapewignercrystalshole,abouelkomsan2026rhombohedralgraphenetalecrystals}, $\epsilon_c(\k)=-\alpha|\k|^2+\beta|\k|^4$, and neglect wave function form factors.
For $\alpha>4\pi n\beta$, the non-interacting Fermi sea is annular, bounded by inner hole-like and outer electron-like Fermi surfaces with radii $k_{F\pm}=\sqrt{(\alpha\pm4\pi n\beta)/(2\beta)}$. In the regime of $k_{F-}\ll k_F\equiv\sqrt{4\pi n}$ and assuming a rotationally invariant liquid, the energy per particle~\eqref{hartree-fock_liquid} reads: 
\begin{equation}
    E/N=E_0/N
    -k_{F-}^2\left(\alpha_*-4\pi n\beta\right)+O(k^3_{F-}),
\end{equation}
where $E_0/N=-\alpha k^2_F/2+\beta k^4_F/3-4\ell k_FE_C/(3\pi)$ is the energy of the simply connected Fermi sea and $\alpha_{*}=\alpha-\ell(\pi-2)E_C/(\pi k_F)$.
To quadratic order, opening an inner hole pocket lowers the energy when $\alpha_*>4\pi n\beta$.
Coulomb exchange reduces $\alpha_*$ relative to its bare value $\alpha$, increasing the energy cost of opening the inner hole pocket and favoring a simply connected Fermi sea.
Fig.~\ref{fig:band_renormalization}(b) shows the line $\alpha_*=4\pi n\beta$, where this quadratic cost vanishes, shifted above the bare Lifshitz line (gray). Quantum geometry, neglected in the Mexican-hat model, can further favor pocket collapse by shifting electrons toward regions of smaller quantum metric, where larger wave function overlaps enhance the exchange-energy gain.

The uniform-liquid analysis excludes a competing possibility: holes in the inner pocket may crystallize while electrons near the outer Fermi surface remain metallic.
This state, recently termed an anticrystal~\cite{abouelkomsan2026rhombohedralgraphenetalecrystals}, is favored at lower electron densities, where the inner hole pocket is more populated. Its formation is favored by the larger effective mass of the hole-like quasiparticles compared with electrons near the outer Fermi surface.
To account for competing states at lower density, we perform large-scale, fully unrestricted HF calculations of the Hamiltonian~\eqref{hamiltonian_ground_state}. 
Details of the method are provided in the SM~\cite{supplementary}; see also Refs.~\cite{Dai2024,nashabeh2026predictionsuperconductivitymassasymmetricelectronhole,abouelkomsan2026rhombohedralgraphenetalecrystals}.
These calculations allow arbitrary real-space density patterns while retaining the nontrivial wave function topology of R$L$G.

Figure~\ref{fig:AC}(a) shows the ground-state charge density $\rho(\r)=A^{-1}\sum_{\k,\p}e^{-i(\k-\p)\cdot\r}\Lambda_{\k,\p}n_{\k,\p}$, obtained from an unrestricted HF with $516$ electrons for $n=0.8\times 10^{12}$cm$^{-2}$ and $u_D=48$meV. Here, $A$ is the system area and $n_{\k,\p}=\langle\hat c^\dagger_{\k}\hat c_{\p}\rangle$ is the one-particle reduced density matrix.
The charge density forms a lattice of minima, each corresponding to a localized hole from the inner Fermi pocket. 
The noninteracting pocket area nearly matches that of the lattice's hexagonal Brillouin zone [Fig.~\ref{fig:AC}(b)], consistent with one hole per unit cell.
The electron filling of this lattice is $\nu=nA_{\rm UC}\approx2.2$, where $A_{\rm UC}=\sqrt{3}a^2/2$ is the area of the unit cell. 
Remarkably, in this regime, the diagonal component $n_{\k,\k}$, shown in Fig.~\ref{fig:AC}(b), exhibits a sharp Fermi-surface discontinuity associated with well-defined electron quasiparticles, supporting the metallic character of this state. 
The discontinuity follows the outer electron-like non-interacting Fermi contour, shown as a solid white line in Fig.~\ref{fig:AC}(b), but is interrupted and reconstructed by the periodic potential of the hole crystal.

Restricted HF calculations independently recover the anticrystal upon imposing a triangular unit cell with lattice vectors $(\a_1,\a_2)$ of equal length and relative angle $2\pi/3$. 
We then minimize the energy by varying the filling factor $\nu=n A_{\rm UC}$, which sets the lattice constant $a=n^{-1/2}(4\nu^2/3)^{1/4}$.
This approach, equivalent to restricting $n_{\k,\p}$ to momentum pairs satisfying $\k-\p=\G$, where $\G$ is a reciprocal lattice vector, resolves the folded band structure shown in Fig.~\ref{fig:AC}(c) and the resulting reconstruction of the electron Fermi surface shown in Fig.~\ref{fig:AC}(d).
A minimum within a restricted variational space does not guarantee stability against more general fluctuations~\cite{Kwan2025,Desrochers2026,Yu2025,Dong_2025,ge2026nematic,Hirsbrunner2026,Zhou2025}. 
Although unrestricted HF calculations recover the anticrystal, its stability against collective fluctuations requires a separate assessment~\cite{Kwan2025,Desrochers2026,Yu2025,Dong_2025,ge2026nematic,Hirsbrunner2026,Zhou2025}. We confirm this stability using time-dependent HF theory~\cite{Thouless1960,Kwan_2025,Wolf_2024,Schuck_2021}, as detailed in the SM~\cite{supplementary}.

\begin{figure}
    \centering
    \includegraphics[width=\linewidth]{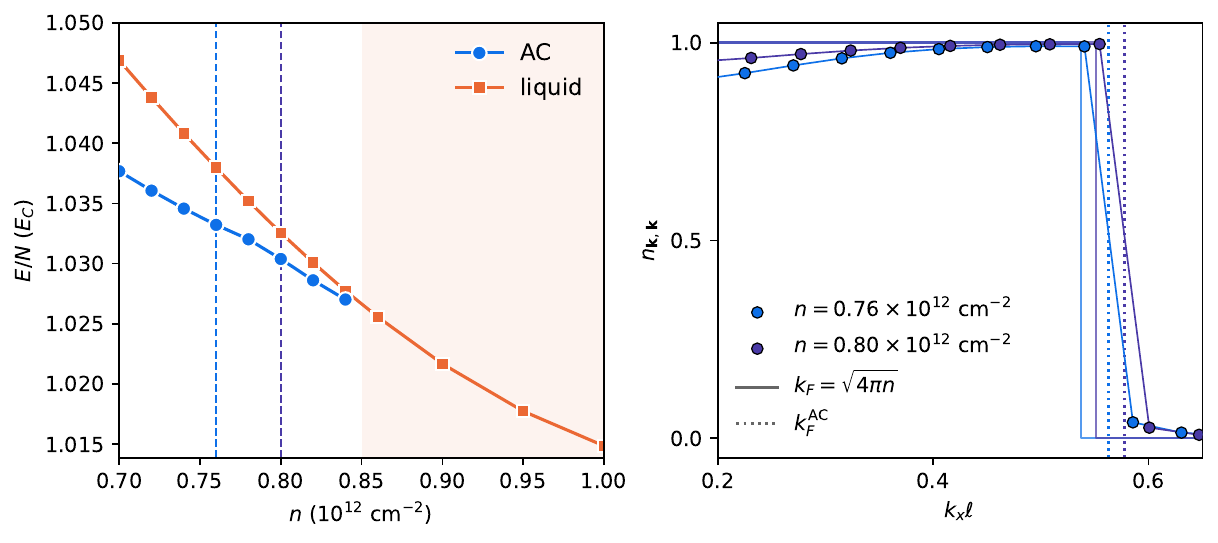}
    \caption{\textbf{Hole crystallization transition and Fermi surface discontinuity:} (a) Energy of the anticrystal (AC) (blue) and of the liquid (orange). (b) Momentum distribution of the metastable liquid (solid lines) and the anticrystal (dots). Results are obtained for $u_D=48$meV, $L=4$, $N=516$ electrons and $\epsilon=15$.}
    \label{fig:FS_jump}
\end{figure}

The energy per particle in Fig.~\ref{fig:FS_jump}(a) shows a transition from the uniform liquid to the anticrystal as the density is lowered, reflecting two competing fates of the inner hole pocket: collapse in the liquid and crystallization in the anticrystal. We now show how this transition produces a jump in the dominant quantum-oscillation frequency, while the periodic potential of the hole crystal generates additional frequencies.

\textit{Quantum oscillations ---} As shown in Fig.~\ref{fig:FS_jump}(a), at $u_D=48$meV the uniform liquid is energetically favored for densities above $n_c\simeq0.85\times10^{12}$cm$^{-2}$. 
In this regime, interactions eliminate the inner hole pocket, leaving a simply connected Fermi sea with a single quantum-oscillation frequency, $F=hn/e$. This frequency is set by the area $A_{\rm FS}=(2\pi)^2n$ enclosed by the interaction-renormalized Fermi surface~\cite{Luttinger1961,Fabrizio2022}. As the density is lowered, both $A_{\rm FS}$ and $F$ decrease, consistent with experimental observations~\cite{Han2025,kalantre2026}.

For $n<n_c$, the anticrystal becomes energetically favored [Fig.~\ref{fig:FS_jump}(a)].
Hole crystallization has two effects: (i) it depletes the electron occupation at low momenta, increasing the outer Fermi wavevector relative to the uniform liquid [Fig.~\ref{fig:FS_jump}(b)]; (ii) it generates a periodic HF potential that folds the bands and reconstructs the Fermi surface [Figs.~\ref{fig:AC}(d)]. 
In this situation, to understand the properties of the oscillations of the longitudinal resistance $R_{xx}$ as function of the magnetic field $B$ magnetic breakdown must be taken into account. 

\begin{figure*}
    \centering
    \includegraphics[width=.9\linewidth]{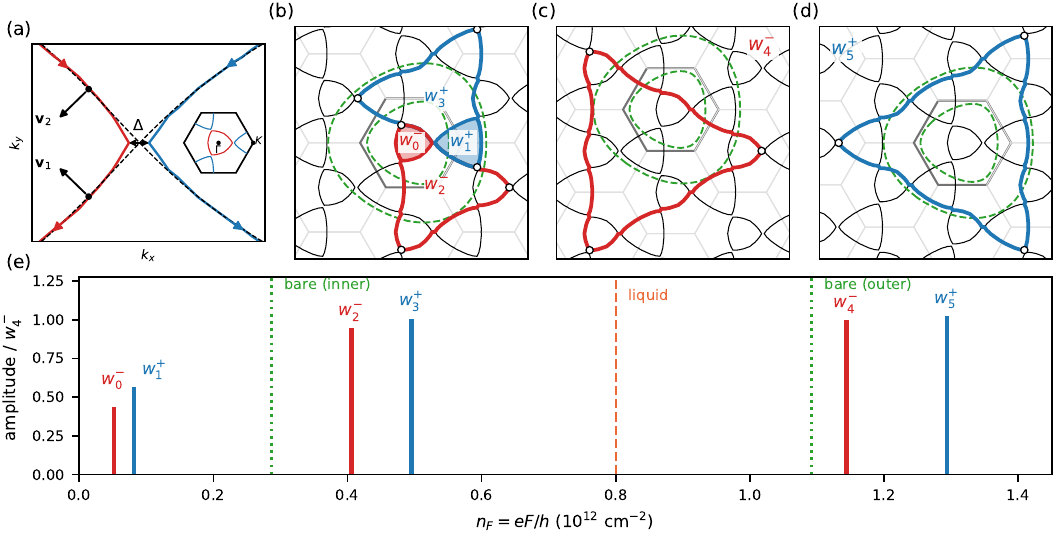}
    \caption{\textbf{Quasiparticle orbits and quantum-oscillation frequencies} ($n=0.8\times10^{12}\,\mathrm{cm^{-2}}$, $u_D=48$meV, $L=4$ and $\epsilon=15$).
(a)~Breakdown junction between the hole pocket at $\Gamma$ (red) and the electron pocket at $K$ (blue); dashed lines are the diabatic trajectories, $\Delta$ the gap.
(b)--(d)~Triangular orbits $w$ on the HF Fermi surface (black): (b)~$w_{0,1,2,3}$; (c)~$w_4$; (d)~$w_5$. Open circles: Bragg reflections; green dashed: non-interacting Fermi surface.
(e)~Amplitudes of the $w$ orbits versus $n_F=eF/h$ at $P=0.92$ ($B=1$\,T); the dashed and dotted lines mark the liquid ($n_F=n$) and the non-interacting Fermi surfaces.}
    \label{fig:orbits}
\end{figure*}

Within the semiclassical approximation and in the limit $\omega_c\tau\gg1$, the oscillatory component of the longitudinal resistance, $\Delta R_{xx}$, is described by the Lifshitz--Kosevich formula~\cite{lifshits1958},
\begin{equation}
 \frac{{\Delta R_{xx}}}{4R_0}=\mathrm{Re}\sum_w\mu^*_w\,W_w(B)\,R_{Tw}\,R_{Dw}\,e^{2\pi i\,\left(\frac{F_w}{B}-\gamma_w\right)}.
 \label{eq.Rxx}
\end{equation}
where $\omega_c$ is the cyclotron frequency, $\tau$ is the electron relaxation time, and $R_0$ is the zero-field resistance. The sum runs over all momentum-space cyclotron orbits $w$ accessible to quasiparticles at the Fermi level.
Magnetic breakdown connects these orbits through tunneling at junctions where distinct Fermi-surface segments approach one another. 
At each junction, quasiparticles either tunnel with probability $P$ or are reflected with probability $Q=1-P$. In our case, all junctions are related by $C_{3z}$ symmetry and therefore share the same tunneling probability,
\begin{equation}\label{eq:probability}
    P=e^{-B_0/B},\qquad
    B_0=\pi\Delta^2/(2e\hbar|\bm v_1\times\bm v_2|),
\end{equation}
where $\Delta$ is the umklapp gap induced by the anticrystal, and $\bm v_1$, $\bm v_2$ are the Fermi velocities of the incoming trajectories shown in Fig.~\ref{fig:orbits}~(a).
From our HF results at $n=0.8\times 10^{12}$cm$^{-2}$, we estimate $\Delta\simeq 1.5\,\mathrm{meV}$ and $|\bm v_1\times\bm v_2|\simeq 6.46\times 10^{1 0}\,\mathrm{m^2/s^{2}}$, yielding $B_0\simeq 0.083$T and $P\simeq 0.92$ at $B=1$T.
%
In Eq.~(\ref{eq.Rxx}), for each orbit $w$, the effect of tunneling and reflection processes 
is taken into account by the breakdown amplitude
\begin{equation}
 W_w(B)=\prod_{\rm crossings}e^{-B_0/2B}\prod_{\rm reflections}i\sqrt{1-e^{-B_0/B}}.
 \label{eq:W}
\end{equation}
$\mu^*_w$ is the ratio between the effective cyclotron mass for a given orbit and a reference mass $m_0$.
In the case in which the Fermi velocities on the different Fermi pockets
are approximately the same, $\mu^*_w$ is simply proportional to the length of the orbit.
Setting $\alpha_0\equiv 2\pi^2 k_B m_0/(\hbar e)$, we can express the thermal factor $R_T$ in
Eq.~(\ref{eq.Rxx}) as 
$
R_{Tw} = \alpha_0\mu^*_w T/[B \sinh(\alpha_0\mu^*_w T/B)]
$
and the Dingle factor $R_D$ as
$
R_{Dw} =  \exp(-\alpha_0\mu^*_w T_{w,D}/B),
$
with $T_{w,D}=\hbar/(2\pi k_B\tau_w)$.
$F_w$ is the Onsager frequency of the signed area $A_w$ enclosed by the orbit $w$,
$F_w=\hbar A_w/(2\pi e)$ and $\gamma_w=1/2-\varphi_{B,w}/2\pi$,
with $\varphi_{B,w}$ the Berry phase for orbit $w$.
Because we are in the QM phase the spin is completely polarized
and so there are no corrections to Eq.~(\ref{eq.Rxx}) arising from
the Zeeman term.

Figures~\ref{fig:orbits}(b)-(d) shows all orbits $w^{\pm}_j$ involving at most three reflections and six crossings on the anticrystal-reconstructed Fermi surface. Red and blue denote hole-like (clockwise) and electron-like (anticlockwise) orbits, respectively. Orbits with $j>0$ connect the pocket around $\Gamma$ to those near $K$ through magnetic breakdown.
\begin{table}[t]
\begin{tabular}{lccccc}
\toprule
orbit &$A_w\,[\ell^{-2}]$ & $F^{\pm}_j/F_l$ & $\quad\mu^*_w\quad$ & $|W_w|$ & amplitude \\
\midrule
$w_0^-$ & 0.056  &  0.059    & 0.16 & $Q^{3/2}$       & 0.42 \\
$w_1^+$ & 0.111  &  0.116    & 0.24 & $Q^{3/2}$       & 0.59 \\
$w_2^-$ & 0.480  &  0.503    & 0.58 & $Q^{3/2}P^{3}$  & 0.94 \\
$w_3^+$ & 0.648  &  0.679    & 0.66 & $Q^{3/2}P^{3}$  & 1.02 \\
$w_4^-$ & 1.385  &  1.452    & 1    & $Q^{3/2}P^{6}$  & 1    \\
$w_5^+$ & 1.664  &  1.744    & 1.08 & $Q^{3/2}P^{6}$  & 1.03 \\
\bottomrule
\end{tabular}
\caption{The superscript in the orbit labels is $-$ for hole-like and $+$ for electron-like circulation. 
For $\mu^*_w$ $m_0$ is set equal to $m^*$ of $w_4^-$.  Amplitude: $|\Delta R_{xx}/R_0|$ relative to $w_4^-$ for $P=0.92$ and $\alpha_0 T/B=\alpha_0 T_{w,D}/B=0.5$; it includes the coherent sum over all walks of equal area, with up to 3 reflections.}
\label{tab.orbits}
\end{table}
Table~\ref{tab.orbits} shows the areas of the orbits, and their amplitudes normalized by the amplitude of the orbit $w_4^-$, shown in red in Fig.~\ref{fig:orbits}(c), that has the closest area to the one of the Fermi surface for $n$ just above $n_c$.
These results assume a common quasiparticle lifetime $\tau_w$ for all orbits.
The relative amplitudes in Table~\ref{tab.orbits} should therefore be compared with experiment only qualitatively. In particular, orbit-dependent relaxation times may account for some of the discrepancies with the amplitudes measured in Ref.~\cite{kalantre2026}.

Despite these approximations, the results in Table~\ref{tab.orbits} capture three key experimental features: 
(i) a dominant frequency $F^-_4$ larger than that associated with the simply connected Fermi surface for $n\gtrsim n_c$, a jump that cannot be explained within a noninteracting picture; (ii) an additional frequency $F^+_5\approx 1.2F^-_4$~\cite{kalantre2026}; and (iii) the emergence of multiple SdH frequencies in the anticrystal regime for $n<n_c$. 

\begin{figure}
    \centering
    \includegraphics[width=.8\linewidth]{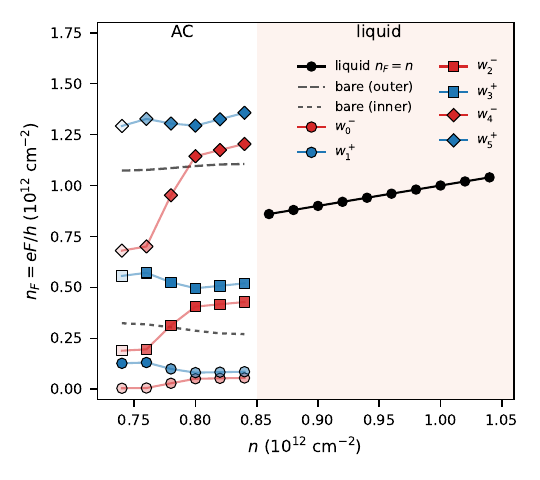}
    \caption{\textbf{Quantum oscillation frequencies across the hole crystallization transition}{ ($u_D=48$meV, $L=4$ and $\epsilon=15$). Different symbols in the anticrystal (AC) regime denote higher electron (blue) and hole (red) orbits, with opacity indicating the oscillation amplitude obtained using Eq.~\eqref{eq.Rxx} where $B_0$~\eqref{eq:probability} and $\mu^*_w$ are computed at each density. 
    In the liquid regime, we find a single frequency from a simply connected Fermi surface.}}
    \label{fig:q_oscillations}
\end{figure}

Extending this analysis across the anticrystal regime in Fig.~\ref{fig:q_oscillations}, we find that the electron-like orbit frequencies depend only weakly on density, whereas the hole-like frequencies decrease substantially below $n\simeq 0.8\times10^{12}\mathrm{cm}^{-2}$. 
This decrease reflects the shrinking hole-like pocket around $\Gamma$ as the electron density is reduced, while the large electron-like orbits remain robust throughout this range. 
Although quantitative agreement with experiment varies with density, our analysis captures the upward jump in the dominant frequency near the transition and produces a multitone SdH spectrum~\cite{kalantre2026}.
These qualitative features suggest that the anticrystal proposed in Ref.~\cite{abouelkomsan2026rhombohedralgraphenetalecrystals} may be relevant to the normal state near the superconducting regions in the QM regime of rhombohedral graphene.

\textit{Discussion ---} The contrasting fates of holes in an annular Fermi sea provide a mechanism for the upward jump in the dominant oscillation frequency~\cite{kalantre2026}: the collapse of the inner hole pocket contracts the liquid’s Fermi surface, whereas the crystallization of the holes preserves a larger outer electron contour.
Magnetic breakdown connects segments of this reconstructed contour into large orbits, as shown in Figs.~\ref{fig:orbits}(c)-(d). 
Near the transition, a prominent hole-like orbit accounts for the upward frequency jump (Fig.~\ref{fig:q_oscillations}), while an additional orbit produces a frequency close to $1.2$ times the dominant one. 
Deeper in the anticrystal regime, the hole-like frequency varies with density, whereas the electron-like frequencies remain comparatively stable.

Real-space imaging of the predicted charge modulation would provide an independent test of the anticrystal interpretation. The corresponding filling $nA_{\mathrm{UC}}$ varies with density and remains close to, but is not pinned at two electrons per unit cell. 
Additionally, the anomalous Hall response observed experimentally in the normal phase is consistent with a metallic anticrystal. 

Our results support the anticrystal as a candidate normal state near superconductivity in \rlg, where the pairing mechanism remains unresolved. Proposed mechanisms include: (i) phonon-mediated pairing in odd-angular-momentum channels, enabled by retardation effects~\cite{Yang-Zhi2026} or by band topology in the static limit~\cite{li2025berrytrashcanshortrange,maymann2025pairingmechanismdictatestopology}; (ii) the Kohn--Luttinger mechanism~\cite{geier2024chiraltopologicalsuperconductivityisospin,Yang-Zhi2025,Yang2025,Guinea2025,Gaggioli2025,Christos2026,Qin2026Chiral,Yoon2026,Patri_2025}, which is enhanced by Berry curvature effects~\cite{Shavit2025,Jahin2026,barlas2025quantumgeometryinducedkekule,Guerci2025,zou2026kohnluttingersuperconductivityflatchern}.
The anticrystal suggests a further possibility: vibrations of the hole crystal could mediate pairing among the itinerant electrons, as proposed for related electronic crystals~\cite{nashabeh2026predictionsuperconductivitymassasymmetricelectronhole,miao2026cooperinstabilitymagneticwigner}. The hole crystal could thus both coexist with and promote superconductivity, opening a route to an electronic supersolid.

%

\textit{Acknowledgments ---} 
We thank Sankar Das Sarma and Andrei Bernevig for insightful discussions. DG thanks Filippo Gaggioli, Ahmed Abouelkomsan, and Liang Fu for fruitful discussions and collaboration on related work. Work by ER was funded by the US Department of Energy, Office of Basic Energy Sciences, via Award DE-SC0022245. 
ER also thanks the Aspen Center for Physics, supported by National Science Foundation Grant No. PHY-2210452, and David Goldhaber-Gordon, whose talk during 
the {\it Topology and Strong Correlations: Emerging Universality} program, 
and subsequent discussion, introduced him to the results of Ref.~\cite{kalantre2026}. 
We acknowledge William \& Mary Research Computing for providing computational resources. 
DG is supported by startup funds at William \& Mary.


\bibliography{biblio}


\onecolumngrid
\newpage
\makeatletter 

\begin{center}
\textbf{\large Supplementary Materials for: ``\@title ''} \\[10pt]
Daniele Guerci$^{1}$ and Enrico Rossi$^{1}$ \\
\textit{$^1$Department of Physics, William \& Mary, Williamsburg, VA-23187, USA}\\
\end{center}
\vspace{10pt}

\setcounter{page}{1}  
\setcounter{figure}{0}
\setcounter{section}{0}
\setcounter{equation}{0}

\renewcommand{\thefigure}{S\@arabic\c@figure}
\makeatother

\appendix

This Supplementary Material describes the microscopic model (Sec.~\ref{app:model}) and the unrestricted and restricted Hartree-Fock calculations (Sec.~\ref{app:HF}). 
We the examine the anticrystal's excitation spectrum and stability using time-dependent Hartree-Fock theory (Sec.~\ref{app:tdhf}). Finally, we develop an analytical description of interaction-driven Fermi-surface reshaping and inner-hole-pocket suppression in the uniform liquid (Sec.\ref{app:inner_hole}).

\section{Microscopic Model for Rhombohedral Multilayer Graphene}\label{app:model}

We begin by presenting the low-energy continuum model for spin and valley-polarized rhombohedral graphene:
\begin{equation}
    H(\r_1,\cdots,\r_N)=\sum_{j=1}^{N} h_K(-i\nabla_j)+\sum_{i<j}V(\r_i-\r_j).
\end{equation}
In the $K$ valley, the Hamiltonian $h_K$ is given by:
\begin{equation}\label{sm:Llayer}
    h_K =
    \begin{pmatrix}
        h^{(0)}_0 & h^{(1)} & h^{(2)} & 0 & \cdots & 0 \\
        h^{(1)\dagger} & h^{(0)}_1 & h^{(1)} & h^{(2)} & \ddots & \vdots \\
        h^{(2)\dagger} & h^{(1)\dagger} & h^{(0)}_2 & h^{(1)} & \ddots & 0 \\
        0 & h^{(2)\dagger} & h^{(1)\dagger} & h^{(0)}_3 & \ddots & h^{(2)} \\
        \vdots & \ddots & \ddots & \ddots & \ddots & h^{(1)} \\
        0 & \cdots & 0 & h^{(2)\dagger} & h^{(1)\dagger} & h^{(0)}_{L-1}
    \end{pmatrix}+h_{\rm ISP},
\end{equation}
where, for simplicity, we dropped the momentum dependency, $u_l =u_D (l-(L-1)/2)$ with $l=0,\cdots,L-1$, $u_D$ the potential energy drop between adjacent layers and: 
\begin{equation}
    h^{(0)}_l(\k) = \begin{pmatrix}
        u_l & -t_0 f_{\k} \\ 
        -t_0 f^*_{\k} & u_l 
    \end{pmatrix},\quad
    h^{(1)}(\k) = \begin{pmatrix}
        t_4 f_{\k} & t_3 f^*_{\k} \\ 
        t_1  & t_4 f_{\k} 
    \end{pmatrix},\quad    h^{(2)} = \begin{pmatrix}
        0 & t_2/2  \\ 
        0 & 0 
    \end{pmatrix},\quad
    f_{\k}=\sum^3_{j=1}e^{i\bm u_j\cdot\k}.
\end{equation}
In the latter expression, we have introduced the vectors connecting the two sublattices $\bm u_j=a_GR^{j-1}_{2\pi/3}(0,1)/\sqrt{3}$, $R_{2\pi/3}$ rotation of $2\pi/3$ around $z$ and $a_G=0.246\,$nm.
These model parameters are consistent with those utilized in Refs.~\cite{Dong_2024,Yu2025}: \(t_0=3.1\,{\rm eV}\), \(t_1=380\,{\rm meV}\), and \((t_2,t_3,t_4)=(-24,290,141)\,{\rm meV}\), while the layer-dependent onsite potential \(u_\ell\) accounts for the applied displacement field.
The model features a characteristic length scale \(\ell=\hbar v_F/t_1=a_G\sqrt{3}t_0/(2t_1) \approx 1.738\,{\rm nm}\), which we use as the unit of length throughout our calculations. 
Following Ref.~\cite{Yu2025,Jonah2024,Kwan2025}, we introduce an inversion-symmetric potential that shifts the local chemical potential of the inner layers relative to the outer layers:
\begin{equation}
    [h_{\rm ISP}]_{l,l'}=V_{\rm ISP}\left|l-\frac{L-1}{2}\right|,
\end{equation}
where $L$ is the number of layers. 
We utilize the value $V_{\rm ISP}=16.65$meV~\cite{Yu2025,Jonah2024,Kwan2025}.  
We introduce the eigenstates and the eigenvalues of the noninteracting Hamiltonian: 
\begin{equation}
   \hat  h(\k) \ket{u_{n}(\k )}=\epsilon_n(\k)\ket{u_{\k n}},  \quad \psi_{\k n,ls}(\r) =\frac{1}{\sqrt{A}}e^{i\k\cdot\r}\braket{l,s}{u_{\k n}}, 
\end{equation}
with $\braket{l,s}{u_{\k n}}$ projection of the Bloch vector in layer space along the component $\ket{l,s}$.

The electron–electron interaction is described by the Coulomb repulsion,
\begin{equation}
H_{\rm int}(\r_1,\cdots,\r_N)=\sum_{i<j}V(\r_i-\r_j),
\end{equation}
where $\r_i$ and $\r_j$ are coordinates in the two-dimensional plane, and we neglect the interlayer separation since the applied displacement field polarizes the conduction-band wave functions toward the outermost layer and the two-dimensional length scales are much larger than the interlayer separation.

We then project the interacting Hamiltonian within the conduction band above charge neutrality: 
\begin{equation}\label{hamiltonian_interacting}
   \hat H = \sum_{\k}\epsilon_c(\k) \hat c^\dagger_{\k} \hat c_{\k} + \frac{1}{2A}\sum_{\q}V(\q):\hat\rho(-\q)\hat \rho(\q):,
\end{equation}
where ``$:\,:$" denotes normal ordering and  we have introduced the the projected density operator: 
\begin{equation}
    \hat \rho(\q) = \sum_{\k} \Lambda_{\k,\k+\q} \hat c^\dagger_{\k}\hat c_{\k+\q},\quad \Lambda_{\k,\k+\q}=\braket{u_{\k c}}{u_{\k+\q c}}. 
\end{equation}
In our calculations, we utilized the long-range Coulomb repulsion: 
\begin{equation}
    V(\q)=\frac{2\pi\ell E_C}{|\q|},\quad     E_C=\frac{e^2}{4\pi\epsilon\epsilon_0\ell}.
\end{equation}
For $\epsilon=15$, the energy is $E_C\approx 55.2$meV.

\subsection{Non-interacting bands}

\begin{figure}
    \centering
    \includegraphics[width=\linewidth]{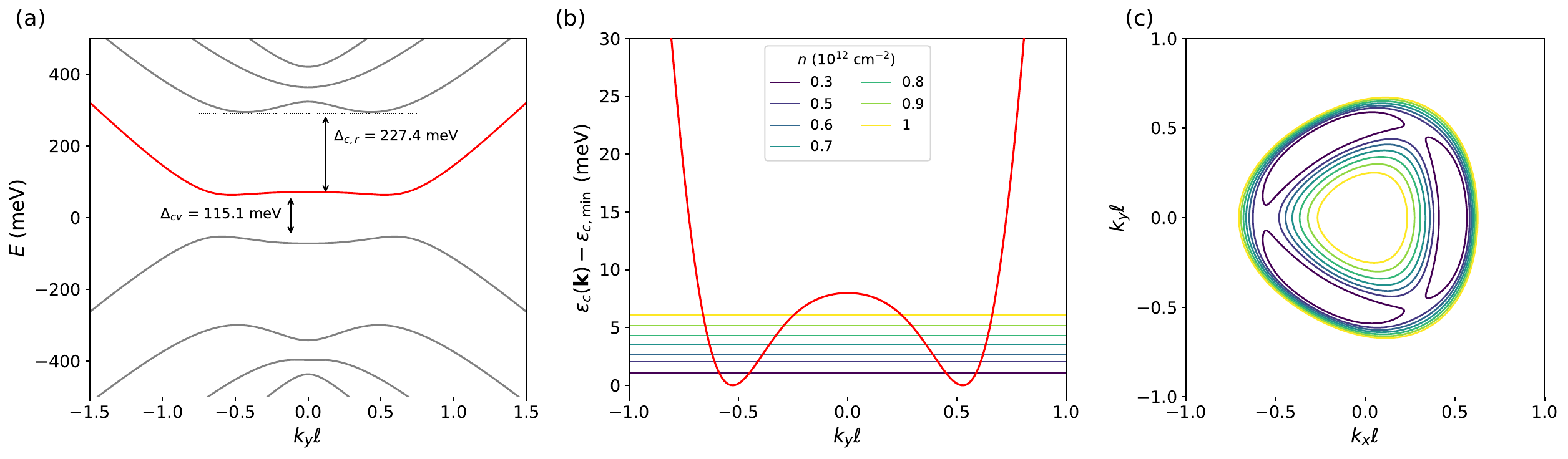}
    \caption{(a) Band structure of rhombohedral tetralayer graphene at $u_D=48\,\mathrm{meV}$; the conduction band above charge neutrality is shown in red. (b) Zoom on this conduction band along $k_y$; horizontal lines mark the Fermi energies for the densities $n$ (in $10^{12}\,\mathrm{cm}^{-2}$) indicated, whose Fermi surfaces are shown in (c) with the same colors.}
    \label{fig:app_bands}
\end{figure}

Fig.~\ref{fig:app_bands}(a) shows the non-interacting bands of tetralayer ($L=4$) rhombohedral graphene for the displacement field $u_D=48$\,meV.
Highlighted in red is the conduction band above charge neutrality, the only band retained in our Hartree--Fock calculations.
It is separated from the valence band by an indirect gap $\Delta_{cv}=115.1$\,meV and from the first remote conduction band by $\Delta_{c,r}=227.4$\,meV $\approx 4.1\,E_C$, where $E_C=e^2/(4\pi\epsilon\epsilon_0\ell)=55.2$\,meV for $\epsilon=15$; projecting onto this single band is therefore well justified at the densities considered.
Near its bottom the band is strongly flattened and takes a ``Mexican-hat'' shape [Fig.~\ref{fig:app_bands}(b)]: the energy at $\k=0$ lies $8.5$\,meV above a trigonally warped ring of minima at $|\k|\ell\simeq0.5$.
As a result, the topology of the non-interacting Fermi sea changes with density as shown in Fig.~\ref{fig:app_bands}(c). 
At $n=0.3\times10^{12}$\,cm$^{-2}$ only the three trigonal minima are occupied, giving three disconnected pockets, which merge into an annulus for $n\gtrsim0.4\times10^{12}$\,cm$^{-2}$.
Up to $n=10^{12}$\,cm$^{-2}$ the Fermi level stays between $1.6$ and $6.6$\,meV above the band minimum, below the central maximum, so the Fermi sea remains annular. 
The kinetic energy per particle is therefore much smaller than $E_{\rm int}=e^2/(4\pi\epsilon_0\epsilon d_{e})$ with $d_e=1/\sqrt{\pi n}$, giving $r_{\rm eff}\equiv E_{\rm int}/\langle\epsilon_{c}(\k)\rangle\gtrsim 20$  for $\epsilon=5$ throughout this range ($r_{\rm eff}\approx 34$ at $n=0.3$ and $r_{\rm eff}\approx 18$ at $n=10^{12}$\,cm$^{-2}$)~\cite{abouelkomsan2026rhombohedralgraphenetalecrystals}.

\section{Hartree-Fock theory}\label{app:HF}

In this section, we provide the Hartree-Fock energy functional for the interacting Hamiltonian~\eqref{hamiltonian_interacting}. 
Introducing the one-particle reduced density matrix (1RDM) $n_{\k,\p}=\mel{\Psi_{\rm HF}}{\hat c^\dagger_{\k}\hat c_{\p}}{\Psi_{\rm HF}}$, the energy functional takes the form: 
\begin{equation}\begin{split}
    E[\hat n] &=\sum_\k\epsilon(\k) n_{\k,\k} + \frac{1}{2A}\sum_{\k_1\cdots\k_4}V_{\k_1\k_2,\k_3\k_4}\left[n_{\k_1,\k_4}n_{\k_2,\k_3}-n_{\k_1,\k_3}n_{\k_2,\k_4}\right]\\
    &=\sum_\k\epsilon(\k) n_{\k,\k} + \frac{1}{A}\sum_{\k_1\cdots\k_4}n_{\k_1,\k_4}n_{\k_2,\k_3} \tilde V_{\k_1\k_2,\k_3\k_4},
\end{split}\end{equation}
where in the last step we have introduced the fully antisymmetric interaction: 
\begin{equation}
    \tilde V_{\k_1\k_2,\k_3\k_4}=\frac{V_{\k_1\k_2,\k_3\k_4}+V_{\k_2\k_1,\k_4\k_3}-V_{\k_1\k_2,\k_4\k_3}-V_{\k_2\k_1,\k_3\k_4}}{4},
\end{equation}
and, in addition, we have 
\begin{equation}
    V_{\k_1\k_2,\k_3\k_4}=\delta_{\k_1+\k_2-\k_3-\k_4,0}V(\k_1-\k_4)\Lambda_{\k_1,\k_4}\Lambda_{\k_2,\k_3}.
\end{equation}

We emphasize that the Hartree-Fock theory is performed on a supercell of sides $\L_1$ and $\L_2$ with periodic boundary conditions and area $A=\L_1\times\L_2$. Correspondingly, the momenta are given by $\k=n_1\Delta\k_1+n_2\Delta\k_2$ where: 
\begin{equation}
       \Delta \k_1= -2\pi \frac{z\times \L_2}{\L_1\times \L_2},\quad    \Delta \k_2= 2\pi \frac{z\times \L_1}{\L_1\times \L_2}.
\end{equation}
Moreover, we introduce an ultraviolet cutoff keeping only momenta $\k$ such that $|\k|<k_{\rm UV}$. 
In the following, the $\q=0$ component of the interaction is excluded. In the direct (Hartree) term, this removes the uniform contribution ($\k_1=\k_4$, $\k_2=\k_3$), which is canceled by the neutralizing background. In the exchange term, it removes the contribution $\propto\sum_{\k_1\k_2}|n_{\k_1,\k_2}|^2=\mathrm{Tr}\hat n^2=\Tr \hat n=N$, which to leading order is a state-independent finite-size shift of order $\sim E_C/\sqrt{A}$ per particle that vanishes as $A\to\infty$. At fixed $N$ and density, $A$ is the same for all fillings, so this energy shift does not affect the comparison between different competing states.

To perform the optimization, we follow Refs.~\cite{Dai2024,nashabeh2026predictionsuperconductivitymassasymmetricelectronhole} in introducing the rectangular matrix $C_{\k,j}$ of  non-orthogonal orbitals with $\k$ run over the momentum grid and $j=1,\cdots,N$ with $N$ number of electrons. 
Under these assumptions, the 1RDM takes the form: 
\begin{equation}
    \hat P=[\hat n]^T= \hat C\cdot \hat S^{-1}\cdot \hat C^\dagger, 
\end{equation}
where we have introduced the matrix $\hat S=\hat C^\dagger \cdot \hat C$. From the definition, we readily find that $n_{\k,\k}\ge 0$ and 
\begin{equation}
    \Tr \hat n=\sum_{\k}\sum_{jj'=1}^{N} C_{\k,j}S^{-1}_{jj'} C^\dagger_{j',\k}=\sum_{j=1}^{N}\left(\hat S^{-1}\cdot \hat S\right)_{jj}=N,\quad \forall\, C_{\k,j}. 
\end{equation}

Within this formulation, the derivative of the energy with respect to $C^*_{\k,j}$ reads: 
\begin{equation}\label{hartree-fock}\begin{split}
    \frac{\delta E}{\delta C_{\p,j}^*}&=\sum_{\k\k'}(\mathbf 1-\hat C\cdot \hat S^{-1}\cdot \hat C^\dagger)_{\p\k} F_{\k,\k'}C_{\k',n}S^{-1}_{nj},\\
    E[\hat n] &= \frac{1}{2}\sum_{\k,\k'}\left[\delta_{\k,\k'}\epsilon(\k) + F_{\k,\k'}\right] n_{\k,\k'}.
\end{split}\end{equation}
$F_{\k,\k'}$ is obtained computing the energy variation with respect to $\hat n$: 
\begin{equation}\begin{split}
   F_{\k,\k'}&= \frac{\delta E[\hat n]}{\delta n_{\k,\k'}}=\delta_{\k,\k'}\epsilon(\k) + \frac{1}{A}\sum_{\p \p'} n_{\p,\p'}\left[V_{\k\p,\p'\k'}-V_{\k\p,\k'\p'}\right].
\end{split}\end{equation}
Hence, given $C_{\k,j}$, we can compute both the gradient and the average value of the energy, which can then be utilized for off-the-shelf minimization. 
At the Hartree-Fock minimum, we have: 
\begin{equation}
    [\hat F,\hat P]=0,\quad \hat P=\hat n^T.
\end{equation}

At self-consistency, we introduce the eigenstates $\eta_j(\k)$ and eigenvalues $E_{j}$ of the Hartree-Fock Hamiltonian: 
\begin{equation}
    \sum_{\k'} F_{\k,\k'}\eta_j(\k')=E_j\eta_j(\k).
\end{equation}
The corresponding Hartree-Fock orbitals of rhombohedral graphene are given by: 
\begin{equation}
    \varphi_j(\x)=\frac{1}{\sqrt{A}}\sum_{\k} e^{i\k\cdot\r}\,\eta_j(\k)\,\braket{l,s}{ u_{\k c}},
\end{equation}
where $\x=(\r,l,s)$ with $l=0,\cdots,L-1$ layer and $s=A,B$ sublattice. 
The Hartree-Fock ground state reads: 
\begin{equation}
    \Psi_{\rm HF}(\x_1,\cdots,\x_N)=\det \varphi_{n}(\x_m),
\end{equation}
where $n,m=1,\cdots,N$ and $\varphi_1,\dots,\varphi_N$ are the $N$ lowest-energy Hartree-Fock orbitals.

\begin{figure}
    \centering
    \includegraphics[width=.9\linewidth]{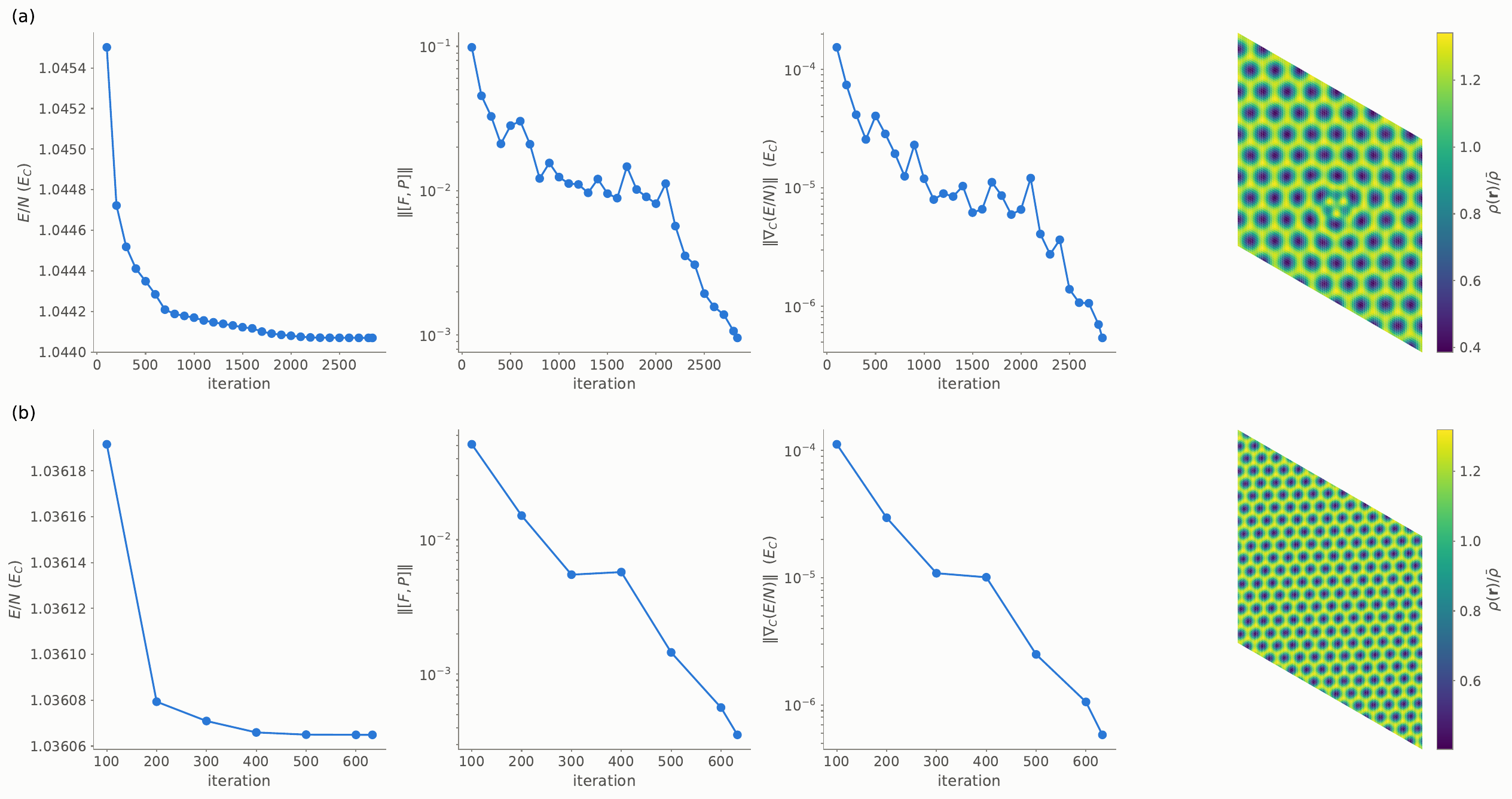}
    \caption{From left to right, the first three panels show the evolution of the energy, the norm of the commutator $[F,P]$, and the gradient of the energy per particle during minimization. The final panel shows the charge distribution of the converged state. The top row (a) corresponds to $192$ electrons, and the bottom row (b) to $516$ electrons. Results obtained for $n=0.72\times10^{12}$cm$^{-2}$, $u_D=48$meV and $\epsilon=15.$}
    \label{fig:hf_minimization}
\end{figure}

\begin{figure}
    \centering
    \includegraphics[width=0.9\linewidth]{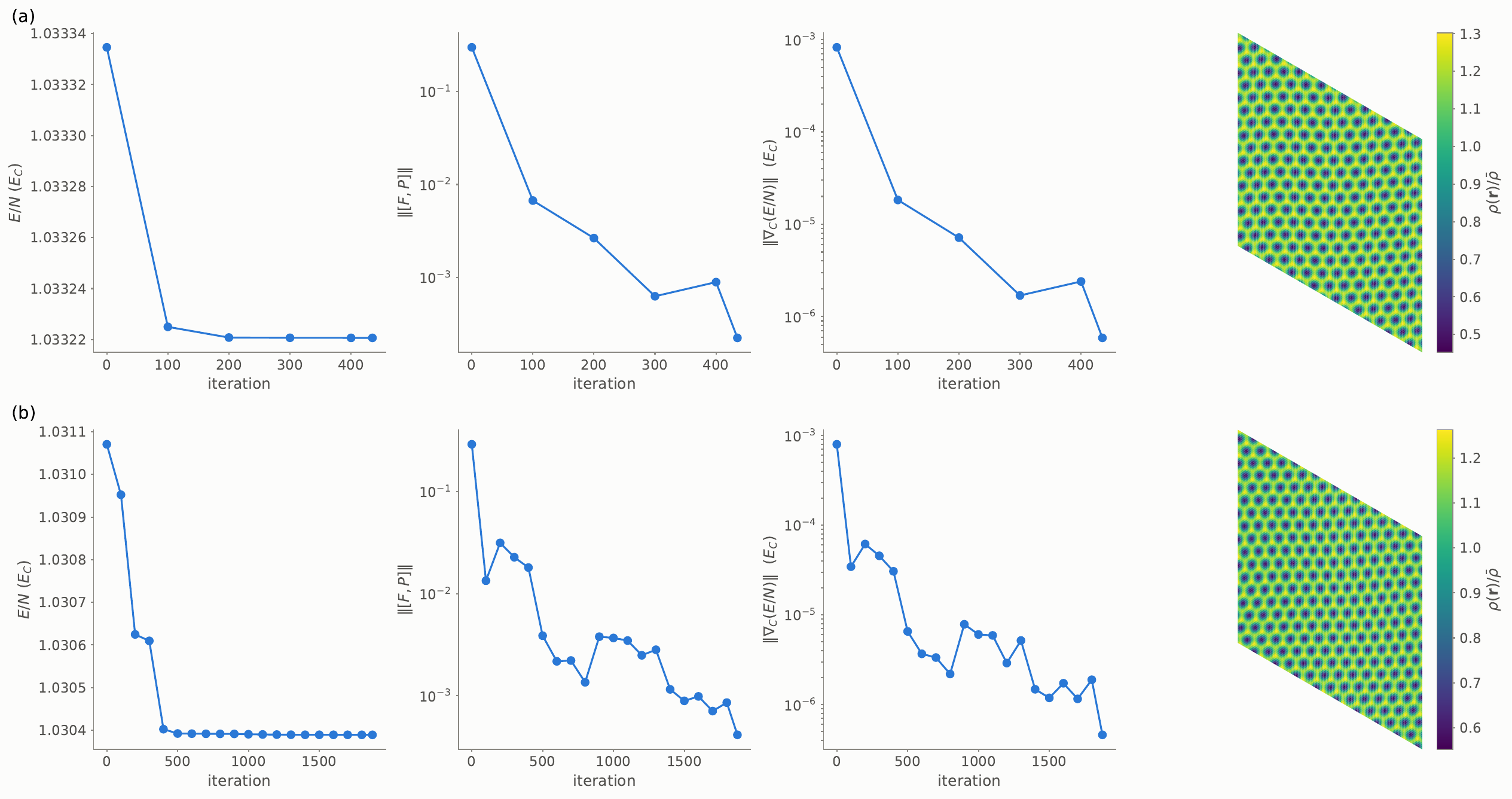}
    \caption{From left to right, the first three panels show the evolution of the energy, the norm of the commutator $[F,P]$, and the gradient of the energy per particle during minimization. The final panel shows the charge distribution of the converged state. The top row (a) corresponds to $n = 0.76 \times 10^{12}\mathrm{cm}^{-2}$, and the bottom row (b) to $n = 0.8 \times 10^{12}\mathrm{cm}^{-2}$ obtained with $516$ electrons. Results obtained for $u_D=48$meV, $L=4$ and $\epsilon=15$.}
    \label{fig:76_8_hf}
\end{figure}

Simulations presented in the main text are performed with $516$ electrons and a varying area $A$ to explore different electron densities $n$. Fig.~\ref{fig:hf_minimization} shows the energy optimization curve for $n=192$ and $n=516$ electrons. 
The top row (a) corresponds to $192$ electrons, initialized from a random 1RDM $n_{\k,\p}$, and the bottom row (b) corresponds to $516$ electrons, obtained by following the anticrystal solution for $n=0.72\times10^{12}$cm$^{-2}$. 
Figure~\ref{fig:76_8_hf} shows the Hartree–Fock minimization at the neighboring densities $n = 0.76 \times 10^{12}\mathrm{cm}^{-2}$ and $n = 0.8 \times 10^{12}\,\mathrm{cm}^{-2}$. The converged solution at $n = 0.72 \times 10^{12}\mathrm{cm}^{-2}$ was used as the initial guess for both calculations.

\subsection{Restricted Hartree-Fock}

In the following, we assume the ground state develops a crystalline order with lattice vectors $(\a_1,\a_2)$. 
Assuming triangular geometry, we consider the lattice vectors 
\begin{equation}
    \a_1=aR_{\theta}(\sqrt{3}/2,-1/2),\quad \a_2=aR_{\theta}(0,1),\quad A_{\rm UC}=\sqrt{3}a^2/2.
\end{equation}
where $\theta$ defines the orientation of the crystal with $\theta=0$ aligned with the graphene crystallographic axis. Correspondingly, we have the reciprocal vectors:
\begin{equation}
    \G_1=-2\pi\frac{\mathbf z\times \a_2}{|\a_1\times\a_2|},\quad \G_2=2\pi\frac{\mathbf z\times \a_1}{|\a_1\times\a_2|}.
\end{equation}
Given the density $n$ and the unit cell $A_{\rm UC}$, we define the filling factor 
\begin{equation}
    \nu=n A_{\rm UC}.
\end{equation}
To find the ground state configuration, we compare different filling factors $\nu$ and minimize the Hartree-Fock energy with respect to the lattice constant: 
\begin{equation}
    a=n^{-1/2}(4\nu^2/3)^{1/4}. 
\end{equation}
Unlike the conventional 2DEG, rhombohedral graphene has additional length scales, such as the inverse radius of the ring of minima underlying the trigonally warped Mexican-hat dispersion in Fig.~\ref{fig:app_bands} and the Berry curvature, making it essential to optimize the crystal periodicity independently of the mean interparticle spacing $d_e=1/\sqrt{\pi n}$.

The number of $\k$ points sampling the Brillouin zone is equal to the number of unit cells:
\begin{equation}
    N_{\k} = \frac{A}{A_{\rm UC}}=\frac{N}{\nu},
\end{equation}
with $\nu$ the filling factor and $N$ the number of electrons.
To compare states at different $\nu$, we keep $N$ fixed and adjust $N_\k$ to maintain a constant supercell area.
We consider supercell with sides: 
\begin{equation}
    \L_1=(s+m)\a_1+m\a_2,\quad \L_2=-m\a_1+s\a_2
    \implies A=\L_1\times\L_2=A_{\rm UC}(s^2+m^2+sm).
\end{equation}
Finally, we emphasize that, for a given density $n$, we fix the number of electrons $N$ and consider the allowed filling factors $\nu$ associated with different choices of $N_{\k}$ and $(s,m)$. 
The energy minimum is found comparing the energy of different Hartree-Fock ground state as a function of $\nu$.

In the folded Brillouin zone, the interaction reads: 
\begin{equation}\label{plane_wave_basis}\begin{split}
    H_{\rm int}=\sum_{[\k_1]\cdots[\k_4]\in\rm BZ}\sum_{\G_1\cdots\G_4}\frac{H^{\G_1\G_2,\G_3\G_4}_{\k_1\k_2,\k_3\k_4}}{2A}\hat c^\dagger_{[\k_1]+\G_1}\hat c^\dagger_{[\k_2]+\G_2}\hat c_{[\k_3]+\G_3}\hat c_{[\k_4]+\G_4},
\end{split}\end{equation}
where $\hat c_{[\k]+\G}$ destroys a plane wave $\psi_{[\k],\G}(\r)=e^{i([\k]+\G)\cdot\r}/\sqrt{A}$ with momentum $[\k]$ in the Brillouin zone and $\G$ the reciprocal lattice vector. 
In the former expression, we have introduced the interaction matrix element: 
\begin{equation}
    H^{\G_1\G_2,\G_3\G_4}_{\k_1\k_2,\k_3\k_4}=V_{[\k_1]+\G_1-[\k_4]-\G_4}\delta_{[\k_4]+[\k_3]-[\k_2]-[\k_1],\Delta\G}\delta_{\G_1+\G_2-\G_3-\G_4,\Delta\G}\Lambda_{[\k_1]+\G_1,[\k_4]+\G_4}\Lambda_{[\k_2]+\G_2,[\k_3]+\G_3},
\end{equation}
with $\Lambda_{\k,\p}$ wave function overlaps introduced previously. 
In addition, we have the kinetic energy term: 
\begin{equation}
    \hat H_{\rm kin}=\sum_{[\k]\in\rm BZ}\sum_{\G}\epsilon([\k]+\G)\hat c^\dagger_{[\k],\G}\hat c_{[\k],\G}.
\end{equation} 
The Hartree term is given by: 
\begin{equation}
    \hat H_H=\sum_{[\k]\in\rm BZ}\sum_{\G\neq \G'}\frac{V(\G-\G')}{A_{\rm UC}}\Lambda_{[\k]+\G,[\k]+\G'}\rho(\G-\G')\hat c^\dagger_{[\k]+\G}\hat c_{[\k]+\G'},
\end{equation}
where the form factor is computed directly for the overlap of the matrix elements. In the latter expression, we have introduced the Fourier transform of the density operator:  
\begin{equation}
    \rho(\Q)=\frac{1}{N_{\k}}\sum_{[\k']\in\rm BZ}\sum_{\G_2}P([\k'])_{\Q+\G_2,\G_2}\Lambda_{[\k']+\G_2,[\k']+\Q+\G_2}.
\end{equation}
Here, we have introduced the projector: 
\begin{equation}
    P([\k])_{\G,\G'}=n([\k])_{\G',\G}=\langle \hat  c^\dagger_{[\k]+\G'} \hat c_{[\k]+\G}\rangle = \sum_{n\in\rm occ}U_{\G,n}([\k])U^*_{\G',n}([\k]).
\end{equation}
For metallic states, it is convenient to introduce a small thermal broadening with a Fermi--Dirac distribution.

The Hartree-Fock Hamiltonian becomes: 
\begin{equation}
    \hat H_{HF}=\sum_{[\k]\in\rm BZ}\sum_{\G\G'}\hat c^\dagger_{[\k]+\G}\left[\delta_{\G,\G'}\epsilon([\k]+\G)+\Sigma^H([\k])_{\G,\G'}+\Sigma^F([\k])_{\G,\G'}\right]\hat c_{[\k]+\G'}.
\end{equation}
where we have introduced the Hartree self-energy: 
\begin{equation}
    \Sigma^H([\k])_{\G,\G'}=\frac{V(\G-\G')}{A_{\rm UC}}\Lambda_{[\k]+\G,[\k]+\G'}\rho(\G-\G').
\end{equation}
The Fock self-energy on the other hand, takes the form: 
\begin{equation}
    \hat H_F=\sum_{[\k]\in\rm BZ}\sum_{\G \G'}\Sigma^F([\k])_{\G,\G'}\hat c^\dagger_{[\k]+\G}\hat c_{[\k]+\G'},
\end{equation}
where we have introduced the Fock self-energy: 
\begin{equation}
    \Sigma^F([\k])_{\G,\G'}=-\frac{1}{A}\sum_{[\k']\in\rm BZ}\sum_{\Q}V_{\k-\k'-\Q}\Lambda_{[\k]+\G,[\k']+\G+\Q}P([\k'])_{\G+\Q,\G'+\Q}\Lambda_{[\k']+\G'+\Q,[\k]+\G'}.
\end{equation}
In the restricted Hartree–Fock calculations we used $N_\G=19$ reciprocal lattice vectors, namely $\G=0$ and the first three shells.

\subsubsection{Numerical results}

\begin{figure}
    \centering
    \includegraphics[width=0.5\linewidth]{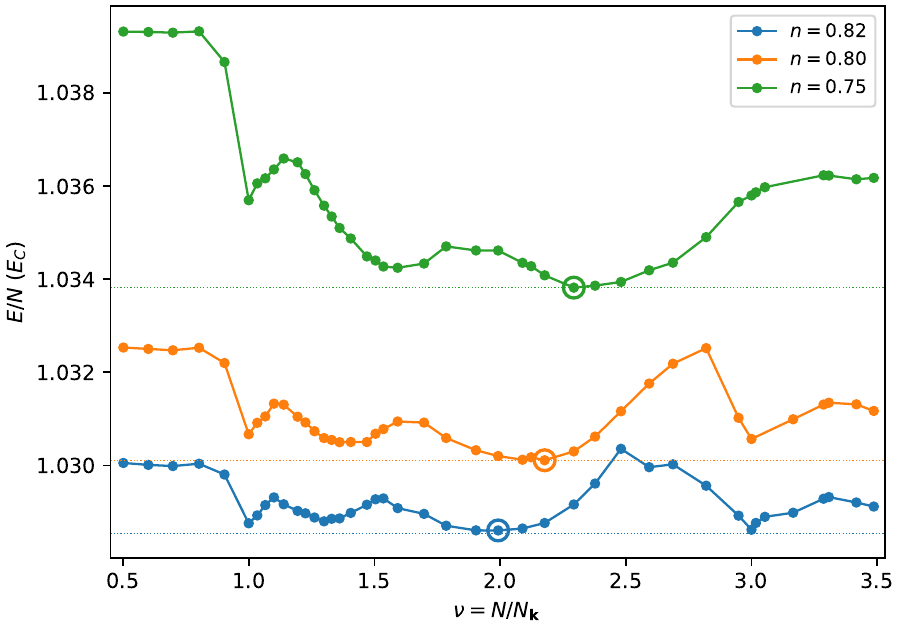}
    \caption{Hartree-Fock energy per particle versus filling $\nu$ for $N=516$, $u_D=48$meV, $L=4$ and $\epsilon=15$.
    Each data point corresponds to the lowest-energy Hartree–Fock solution obtained from six independent initial seeds.}
    \label{fig:hf_energy}
\end{figure}

In this section, we summarize the restricted Hartree–Fock results. 
Figure~\ref{fig:hf_energy} shows the energy per particle as a function of the filling factor $\nu$ at three densities, $n = 0.75$, $0.80$, and $0.82 \times 10^{12}\mathrm{cm}^{-2}$. 
At each density, the lowest-energy state is a metallic anticrystal. The energy minimum is not pinned to a commensurate filling factor, consistent with the metallic nature of this state~\cite{abouelkomsan2026rhombohedralgraphenetalecrystals}. 

Besides the global minimum, the energy $E/N$ displays a second, shallower minimum at lower filling, $\nu \simeq 1.3$–$1.55$ (shifting to larger $\nu$ as $n$ decreases). The corresponding metallic state is a unidirectional stripe: its charge modulation involves a single pair of wavevectors $\pm\bQ$. The stripe lies only $0.2$–$0.45\mathrm{m}E_C$ ($E_C=55.2$meV) energy per particle above the anticrystal minimum, with the smallest gap at the highest density, so the two orders compete closely. 
At $\nu = 3$ for $n = 0.80$ and $0.82$, $E/N$ shows a sharp dip: there the electrons form a commensurate, honeycomb crystal with one electron per $\sqrt{3}$-reduced cell, rotated by $30^\circ$ relative to the imposed lattice, which at $n = 0.82$ is almost degenerate with the metallic anticrystal within $0.08\mathrm{m}E_C$.
Finally, at $\nu = 1$, the commensurate state evolves continuously from a weakly modulated, nearly gapless crystal at $n \gtrsim 0.8\times10^{12}\mathrm{cm^{-2}}$ to a strongly modulated, gapped Wigner-like crystal (gap $\approx 4$ meV) at $n = 0.60$–$0.65\times10^{12}\mathrm{cm^{-2}}$, but it remains a metastable local minimum, $\sim 0.2$–$9\,\mathrm{m}E_C$ above the lowest energy state.

\begin{figure}
    \centering
    \includegraphics[width=0.8\linewidth]{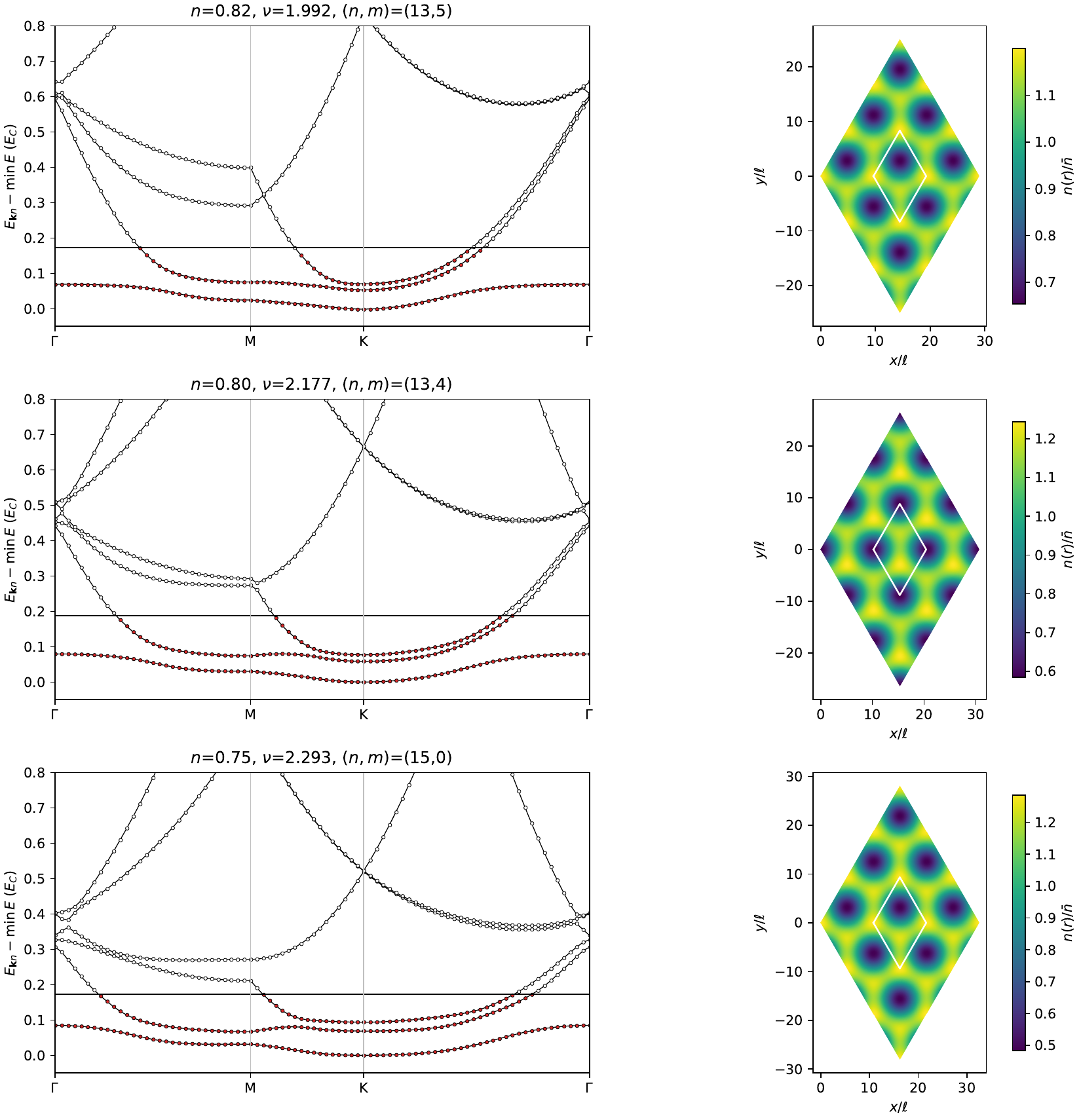}
    \caption{Band structures and charge densities obtained for the states minimizing the energy of restricted Hartree-Fock calculations at $u_D=48$meV, $L=4$ and $\epsilon=15$. The energy unit is $E_C\approx 55.2$meV. }
    \label{fig:hf_bands}
\end{figure}

The corresponding band structures and charge densities are shown in Fig.~\ref{fig:hf_bands}. At the Fermi level, an electron pocket appears around $K$ and a hole pocket around $\Gamma$, similar to the Fermi surface shown in the main text. Reducing the density decreases the size of the hole pocket and, correspondingly, the quantum oscillation frequency associated with hole-like orbits.
Finally, we note that the anticrystal is oriented parallel to the underlying graphene lattice: its primitive vectors are aligned with the graphene zigzag directions, so that the mini-Brillouin-zone corners lie along the graphene valley $K$.

\section{Time-dependent Hartree-Fock}\label{app:tdhf}

To confirm the stability of metallic anticrystals, we perform time-dependent Hartree-Fock calculations.  To this aim, we first express the our interacting Hamiltonian~\eqref{hamiltonian_interacting} in the Hartree-Fock basis $\hat\gamma_{[\k] n}$: 
\begin{equation}\label{hf_basis}
    \hat c_{[\k]+\Q}=\sum_n U_{\Q, n}([\k])\hat \gamma_{[\k] n},
\end{equation}
with $\hat\gamma_{[\k] n}$ destroying an electron in the Hartree-Fock Bloch wave function: 
  \begin{equation}
    \psi_{[\k] n}(\x)=\frac{1}{\sqrt{A}}e^{i[\k]\cdot\r}\sum_{\G}e^{i\G\cdot\r}U_{\G,n}([\k])\braket{l,s}{u_{[\k]+\G c}}.
\end{equation}
By definition, the Hartree-Fock basis diagonalizes the Hartree-Fock Hamiltonian: 
\begin{equation}
    H^{\rm HF}[P]=\sum_{[\k]}\epsilon^{\rm HF}_{[\k]n} \hat \gamma^\dagger_{[\k]n} \hat \gamma_{[\k]n},
\end{equation}
where $\epsilon^{\rm HF}_{[\k]n}$ is the Hartree-Fock dispersion. 

\begin{figure}
    \centering
    \includegraphics[width=.8\linewidth]{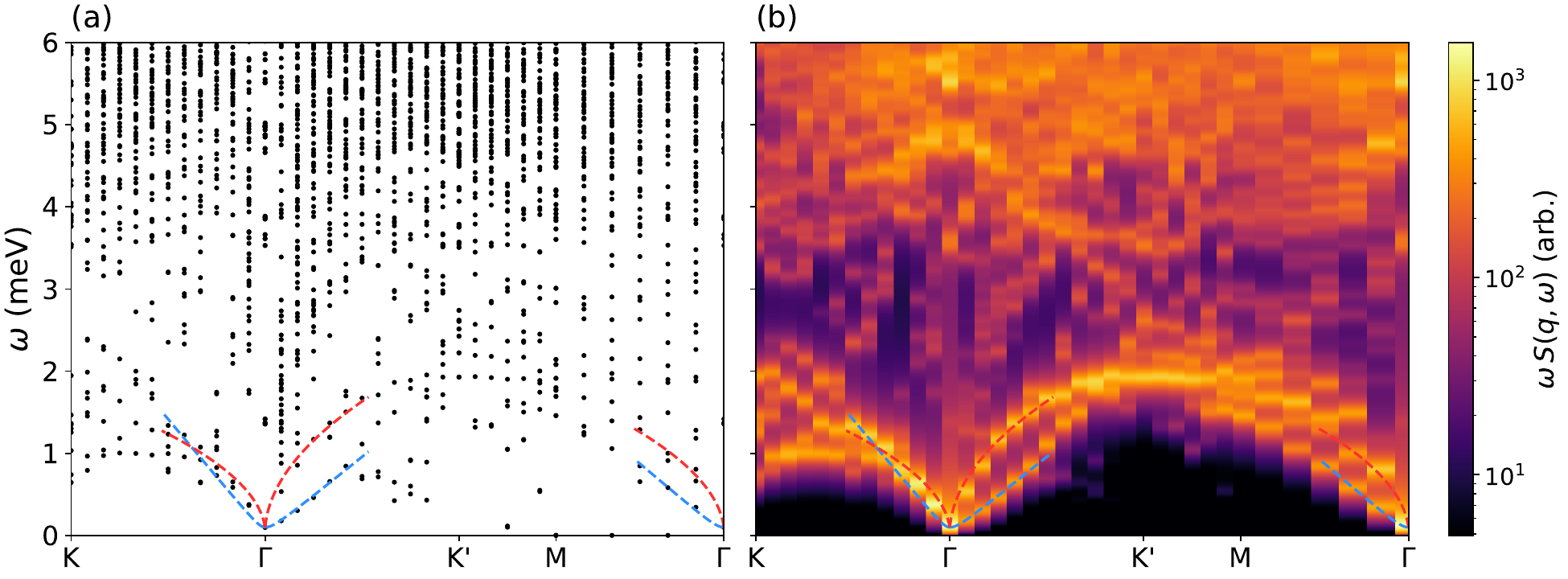}
    \caption{Collective modes of the metallic anticrystal computed with time-dependent Hartree-Fock. Calculations are performed for $N=900$ at $u_D=48$meV, $L=4$, $n=0.8\times10^{12}$cm$^{-2}$ and $\epsilon=15$.}
    \label{fig:collective_modes}
\end{figure}

In this basis, the interaction Hamiltonian~\eqref{plane_wave_basis} is further projected in the self-consistent Hartree-Fock basis~\eqref{hf_basis} and reads:
\begin{equation}
    H_{\rm int}=\frac{1}{2}\sum_{[\k_1]\cdots[\k_4]}\sum_{n_1\cdots n_4}H^{n_1n_2,n_3n_4}_{[\k_1][\k_2],[\k_3][\k_4]}\hat \gamma^\dagger_{[\k_1]n_1}\hat \gamma^\dagger_{[\k_2]n_2}\hat \gamma_{[\k_3]n_3}\hat \gamma_{[\k_4]n_4},
\end{equation}
where we have introduced the projected matrix element: 
\begin{equation}\label{interaction_matrix_element}
    H^{n_1n_2,n_3n_4}_{[\k_1][\k_2],[\k_3][\k_4]}=\frac{1}{A}\sum_{\Q} \delta_{[\k_1]+[\k_2]-[\k_3]-[\k_4],\Delta G}V_{[\k_1]-[\k_4]-\Q}\tilde\Lambda^{n_1,n_4}_{[\k_1],[\k_4]+\Q}\tilde\Lambda^{n_2,n_3}_{[\k_2],[\k_3]-\Q+\Delta\G}.
\end{equation}
In the latter expression, we have introduced the form factors obtained from the Hartree-Fock orbitals: 
\begin{equation}
    \tilde \Lambda^{n,n'}_{[\k],[\k']+\Q}=\int_{\rm UC}\frac{d^2\r}{|\a_1\times\a_2|}e^{-i\Q\cdot\r}U^\dagger_{[\k]n}(\r) U_{[\k']n'}(\r)=\sum_{\G}U^*_{\G,n}([\k])U_{\G+\Q,n'}([\k'])\braket{u_{[\k]+\G c}}{u_{[\k']+\G+\Q c}},
\end{equation}
where $U_{[\k]n}(\r)$ is the cell-periodic wave $U_{[\k]n}(\r)=\sqrt{A}e^{-i\k\cdot\r}\psi_{[\k]n}(\r)$. 
Moreover, we have the kinetic energy term: 
\begin{equation}
    H_{\rm kin} =\sum_{[\k]}\sum_{nn'}\hat \gamma^\dagger_{[\k]n}T_{nn'}([\k])\hat \gamma_{[\k]n'},
\end{equation}
where we have introduced the kinetic energy matrix: 
\begin{equation}
    T_{nn'}([\k])=\sum_{\G}\left[U^\dagger([\k])\right]_{n,\G}\epsilon([\k]+\G)U_{\G,n'}([\k]).
\end{equation}

The spectrum of collective modes is obtained by diagonalizing the dynamical matrix~\cite{Thouless1960}: 
\begin{equation}
    \begin{pmatrix}
        \mathcal A & \mathcal B \\
        -\mathcal B^* & -\mathcal A^*
    \end{pmatrix}\begin{pmatrix}
        X\\
        Y
    \end{pmatrix} =\omega \begin{pmatrix}
        X\\
        Y
    \end{pmatrix},
\end{equation}
where we have introduced the matrix elements: 
\begin{equation}\begin{split}
    \mathcal A_{([\k_1]n_1,[\k_2]n_2);([\k_3]n_3,[\k_4]n_4)}&=\delta_{[\k_1],[\k_3]}\delta_{n_1,n_3}\delta_{[\k_2],[\k_4]}\delta_{n_2,n_4}(\epsilon^{\rm HF}_{[\k_1]n_1}-\epsilon^{\rm HF}_{[\k_2]n_2}) + H^{n_1n_4,n_3n_2}_{[\k_1][\k_4],[\k_3][\k_2]}-H^{n_1n_4,n_2n_3}_{[\k_1][\k_4],[\k_2][\k_3]},\\
    \mathcal B_{([\k_1]n_1,[\k_2]n_2);([\k_3]n_3,[\k_4]n_4)}&=H^{n_1n_3,n_4n_2}_{[\k_1][\k_3];[\k_4][\k_2]}-H^{n_1n_3,n_2n_4}_{[\k_1][\k_3];[\k_2][\k_4]},
\end{split}\end{equation}
where $[\k_1]n_1$ and $[\k_3]n_3$ are empty, while $[\k_2]n_2$ and $[\k_4]n_4$ are occupied. 
Exploiting momentum conservation, we introduce the momentum transfer $[\q]=[\k_1]-[\k_2]=[\k_3]-[\k_4]\in\rm BZ$, and setting $\k_1=\k$ and $\k_3=\p$, we have: 
\begin{equation}\begin{split}
    \mathcal  A_{([\k]n_1n_2);([\p]n_3n_4)}([\q])=\delta_{n_1,n_3}\delta_{n_2,n_4}\delta_{[\k],[\p]}(\epsilon^{\rm HF}_{[\k]n_1}-\epsilon^{\rm HF}_{[\k]-[\q]n_2})+H^{n_1n_4,n_3n_2}_{[\k][\p]-[\q],[\p][\k-\q]}-H^{n_1n_4,n_2n_3}_{[\k][\p]-[\q],[\k-\q][\p]}.
\end{split}\end{equation}
For $B$ we have
\begin{equation}\begin{split}
    \mathcal  B_{([\k]n_1n_2);([\p]n_3n_4)}([\q])=H^{n_1n_3,n_4n_2}_{[\k][\p],[\p+\q][\k-\q]}-H^{n_1n_3,n_2n_4}_{[\k][\p],[\k-\q][\p+\q]}.
\end{split}\end{equation}
For a given collective crystal momentum $[\mathbf q]$, the matrices $\mathcal  A$ and $\mathcal B$ have size $N^2_{\k}\, N^2_e\, N^2_o$ with $N_{\mathrm e}$ and $N_{\mathrm o}$ the numbers of retained empty and occupied bands, respectively. 
With these momentum-resolved definitions, the dynamical equation reads: 
\begin{equation}
    \begin{pmatrix}
        \mathcal A([\q]) & \mathcal B ([\q]) \\
        -\mathcal B^*(-[\q]) & -\mathcal A^*(-[\q])
    \end{pmatrix}\begin{pmatrix}
        X^{[\q]}\\
        Y^{[\q]}
    \end{pmatrix} =\omega_{[\q]} \begin{pmatrix}
        X^{[\q]}\\
        Y^{[\q]}
    \end{pmatrix}.
\end{equation}
The corresponding eigenvectors determine the bosonic operator
\begin{equation}
\hat \beta_{[\q]\nu}^\dagger
=
\sum_{[\k]n_{\rm e}n_{\rm o}}
\left[
X_{[\k]n_{\rm e}n_{\rm o}}^{[\q]\nu}
\hat\gamma_{[\k]n_{\rm e}}^\dagger
\hat\gamma_{[\k-\q]n_{\rm o}}
-
Y_{[\k]n_{\rm e}n_{\rm o}}^{[\q]\nu}
\hat\gamma_{[\k+\q]n_{\rm o}}^\dagger
\hat\gamma_{[\k]n_{\rm e}}
\right],
\end{equation}
where $([\k]n_{\rm e})$ and $([\k\pm \q]n_{\rm o})$ label empty and occupied Hartree--Fock bands, respectively. Both terms transfer crystal momentum $[\q]$. Within the quasiboson approximation, the operators obey bosonic commutation relations when the eigenvectors satisfy
\begin{equation}
\sum_{[\k]n_{\rm e}n_{\rm o}}
\left[
\left(X_{[\k]n_{\rm e}n_{\rm o}}^{[\q]\nu}\right)^*
X_{[\k]n_{\rm e}n_{\rm o}}^{[\q]\mu}
-
\left(Y_{[\k]n_{\rm e}n_{\rm o}}^{[\q]\nu}\right)^*
Y_{[\k]n_{\rm e}n_{\rm o}}^{[\q]\mu}
\right]
=\delta_{\nu\mu}.
\end{equation}

\subsubsection{Spectrum of excitations of the metallic anticrystal}

Fig.~\ref{fig:collective_modes}(a) shows the charge-neutral excitation spectrum of the anticrystal, including collective modes of the electronic crystal and low-energy particle-hole excitations associated with its Fermi surface. 
The longitudinal and transverse modes have real frequencies, with $\omega_L\propto\sqrt{|\q|}$ and $\omega_T\propto|\q|$, respectively.  
Their dispersions are anisotropic, reflecting the broken inversion symmetry.
Moreover, We find that the crystal is mechanically stable, with positive stiffness and real eigenvalues of the dynamical matrix. 

To highlight modes that couple strongly to density fluctuations, Fig.~\ref{fig:collective_modes}(b) shows the frequency-weighted dynamical structure factor $\omega S([\q],\omega)$, where
\begin{equation}
     S([\q],\omega)=\frac{\pi}{A}\sum_{n>0}\sum_{\Q}
    \left|\mel{\Psi_n}{\hat{\rho}_{-[\q]-\Q}}{\Psi_0}\right|^2
    \delta(\omega-\omega_n)\simeq\frac{\pi}{A}\sum_{\nu}\sum_{\Q}
    \left|\mel{\Psi_{[\q]\nu}}{\hat{\rho}_{-[\q]-\Q}}{\Psi_0}\right|^2
    \delta(\omega-\omega_{[\q]\nu}).
\end{equation}
Here, $[\q]$ lies in the crystal Brillouin zone, and $\Q$ runs over the retained reciprocal-lattice vectors. 
The delta functions are broadened into Lorentzians for visualization. The excited states are constructed 
\begin{equation}
    \ket{\Psi_{[\q]\nu}}=\hat\beta^\dagger_{[\q]\nu}\ket{\Psi_0},\quad \hat\beta_{[\q]\nu}\ket{\Psi_0}=0,
\end{equation}
where $\ket{\Psi_0}$ is the RPA ground state~\cite{Fabrizio2022}. The matrix element of the density operator is given by: 
\begin{equation}\begin{split}
    \mel{\Psi_{[\q]\nu}}{\hat{\rho}_{-[\q]-\Q}}{\Psi_0}&=\mel{\Psi_{0}}{[\hat\beta_{[\q]\nu},\hat{\rho}_{-[\q]-\Q}]}{\Psi_0}\\
    &=\sum_{[\k] n_{\rm e}n_{\rm o}}\left[{X^{[\q]\nu*}_{[\k]n_{\rm e}n_{\rm o}}}\tilde \Lambda^{n_{\rm e}n_{\rm o}}_{[\k]+\Q,[\k]-[\q]}+{Y^{[\q]\nu*}_{[\k]n_{\rm e}n_{\rm o}}}\tilde \Lambda^{n_{\rm o}n_{\rm e}}_{[[\k]+[\q]]+\Q+\Delta\G,[\k]}\right],
\end{split}\end{equation}
where we have introduced $[\k]+[\q]-[[\k]+[\q]]=\Delta\G$.

\begin{figure}
    \centering
    \includegraphics[width=0.35\linewidth]{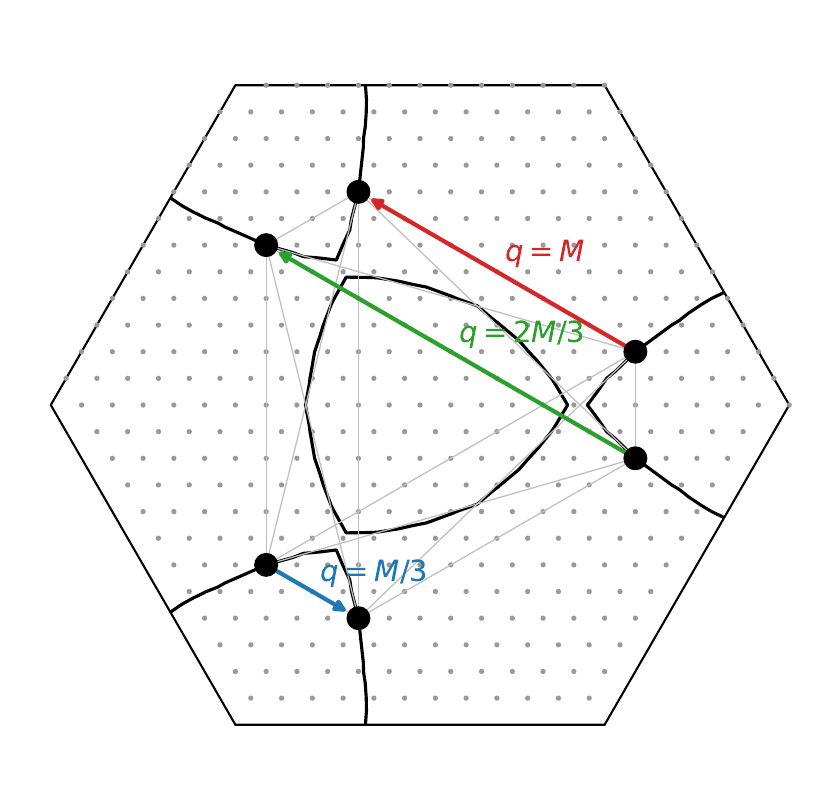}
    \caption{On-shell particle-hole processes. Gray dots indicate the $\k$ points sampling the Brillouin zones, while black dots highlight the open shell involved in particle-hole transitions on the Fermi surface (black line) with center of mass momentum $M$, $2M/3$ and $M/3$.  
    Calculations are performed for $N=900$ (restricted) at $u_D=48$meV, $n=0.8\times10^{12}$cm$^{-2}$ and $\epsilon=15$.}
    \label{fig:on_shell}
\end{figure}

Finally, in addition to collective modes, the spectrum contains low-energy particle-hole excitations associated with the anticrystal’s open-shell Fermi surface. These appear as nearly zero-energy modes along $M$–$\Gamma$, where the transferred momentum connects degenerate, partially occupied states at the chemical potential. Figure~\ref{fig:on_shell} illustrates these processes: gray points mark the sampled $\k$ points, while black dots highlight six degenerate states connected by momenta $M$, $2M/3$, and $M/3$, allowing particle-hole excitations at vanishing energy cost.

\section{Fermi surface renormalization in the annular regime}\label{app:inner_hole}

Luttinger's theorem relates the carrier density to the momentum-space area occupied by a Fermi sea. For a rotationally invariant, spin- and valley-polarized metal with a simply connected Fermi sea, this fixes the Fermi wavevector $k_F$. 
A Mexican-hat dispersion, however, can support an annular Fermi sea bounded by inner and outer Fermi contours. In this case, the density constrains only the difference between their enclosed areas,
\begin{equation}
n=\frac{k_{F+}^2-k_{F-}^2}{4\pi},
\end{equation}
allowing interactions to shift the individual Fermi momenta $k_{F+}$ and $k_{F-}$ while preserving this difference.
In the following, we measure momenta in units of $\ell^{-1}$ and density in units of $\ell^{-2}$, so that $\alpha$ and $\beta$ have units of energy.

At a fixed density $n$, the Hartree-Fock dispersion of the liquid state reads: 
\begin{equation}\label{app:hf}
    \epsilon^{HF}_c(\k)=\epsilon_c(\k)-\frac{1}{A}\sum_{\k'\neq \k}V(\k-\k')|\Lambda_{\k,\k'}|^2 n_{\k'},
\end{equation}
where $n_{\k}=\langle c^\dagger_{\k} c_{\k} \rangle$ is the occupation number. 
In the liquid like states, we observe numerically the tendency of the interaction to remove the inner pocket in the Mexican hat regime.
The numerical analysis is performed minimizing the energy: 
\begin{equation}
    \epsilon[\hat n]=\frac{E[\hat n]}{N}=\frac{1}{n}\int \frac{d^2\k}{(2\pi)^2}n_{\k} \epsilon_c(\k)-\frac{1}{2n}\int \frac{d^2\k d^2\k'}{(2\pi)^4}V(\k-\k')|\Lambda_{\k,\k'}|^2 n_{\k}n_{\k'},
\end{equation}
where $n$ is the electric density. 

To understand this effect, we consider a simplified conduction band dispersion $\epsilon_c(\k)=-\alpha k^2+\beta k^4$ and, for the moment, neglect the form factors $\Lambda_{\k,\k'}=1$. 
In the following, we assume the Fermi liquid state preserves rotational symmetry. 
Given $k_{F-}$ and $k_{F+}$, we introduce the Fock self-energy:  
\begin{equation}
    \Sigma^{F}(\k)=-\frac{1}{A}\sum_{\k'\neq \k}V(\k-\k') n_{\k'}=-\int\frac{d^2\k'}{(2\pi)^2}\frac{2\pi E_C}{|\k-\k'|}n_{\k'}=-E_C\int_{k_{F-}}^{k_{F+}}\int^{2\pi}_0\frac{d\phi}{(2\pi)} \frac{dp\, p}{\sqrt{k^2+p^2-2kp\cos\phi}},
\end{equation}
where we assumed an annular Fermi surface with inner and outer momenta $k_{F-}$ and $k_{F+}$. We observe that $k_{F-}$ and $k_{F+}$ are related to the electron density by the relation: 
\begin{equation}
    n=\frac{k^2_{F+}-k^2_{F-}}{4\pi},
\end{equation}
with $k_{F_+}> k_{F-}$ and $k_{F-}\ge0$.
We notice that the annular Fermi surface regime is realized when $\alpha-4\pi\beta n>0\implies n<\alpha/(4\pi \beta)$. The energy reads: 
\begin{equation}
    \epsilon[\hat n]=\frac{1}{n}\int_{k_{F-}}^{k_{F+}}\frac{ k dk }{2\pi} (-\alpha k^2+\beta k^4)-\frac{E_C}{4\pi n}\int^{2\pi}_0\frac{d\phi}{2\pi}\int_{k_{F-}}^{k_{F+}}\int_{k_{F-}}^{k_{F+}}dk\, dk'\frac{kk'}{\sqrt{k^2+k'^2-2kk'\cos\phi}}.
\end{equation}
Given $n$, the variational energy is minimized with respect to $k_-$. 
The kinetic energy contribution to the energy is given by: 
\begin{equation}
    \epsilon_{\rm kin}[\hat n]=-\alpha k^2_{F-}+\beta k^4_{F-}-2\pi \alpha n+4\pi n\beta k^2_{F-}+\frac{16\pi^2}{3}\beta n^2.
\end{equation}
In the absence of interactions, the value of $k_{F-}$ can be found minimizing the kinetic energy, setting $x=k^2_{F-}$ one obtains: 
\begin{equation}
    \frac{d\epsilon_{\rm kin}}{dx}=-\alpha+2\beta x +4\pi n\beta=0\to k_{F\pm}=\sqrt{\frac{\alpha\pm4\pi n\beta}{2\beta}},
\end{equation}
that is realized when $n<\alpha/(4\pi\beta)$.

The interacting energy, on the other hand, reads: 
\begin{equation}
    \epsilon_{\rm int}[\hat n]=-\frac{E_C}{4\pi n}\int^{2\pi}_0\frac{d\phi}{2\pi}\int_{k_{F-}}^{k_{F+}}\int_{k_{F-}}^{k_{F+}}dk\, dk'\frac{kk'}{\sqrt{k^2+k'^2-2kk'\cos\phi}},
\end{equation}
which can be expressed in terms of elliptic functions. Specifically, utilizing the result: 
\begin{equation}
    \int^{2\pi}_0\frac{d\phi}{2\pi}\frac{1}{\sqrt{k^2+k'^2-2kk'\cos\phi}}=\frac{2}{\pi(k+k')}K\left(\frac{4kk'}{(k+k')^2}\right),
\end{equation}
where $K(x)$ is the complete elliptic integral with parameter $x$. We find: 
\begin{equation}
    \epsilon_{\rm int}[\hat n]=-\frac{E_C}{2\pi^2 n}\int^{k_{F+}}_{k_{F-}}dk\int^{k_{F+}}_{k_{F-}}dk'\,\frac{kk'}{k+k'}K\left(\frac{4kk'}{(k+k')^2}\right).
\end{equation}
Moreover, the self-energy reads: 
\begin{equation}
    \Sigma^{F}(k)=-\frac{2E_C}{\pi}\int^{k_{F+}}_{k_{F-}}dk'  \frac{k'}{k+k'}K\left(\frac{4kk'}{(k+k')^2}\right),
\end{equation}
where $\Sigma^F$ depends only on the absolute value $k=|\k|$.
To gain insight on the effect of the interaction, it is useful to consider the regime of small $k_{F-}$ just below the Lifshitz transition where the hole-like Fermi surface appears. 
We have: 
\begin{equation}
    \frac{d\epsilon_{\rm int}[\hat n]}{dk_{F-}}=\frac{k_{F-}}{2\pi n}\left[\Sigma^F(k_{F+})-\Sigma^F(k_{F-})\right]\implies \epsilon_{\rm int}[\hat n]=-\frac{4 E_C \sqrt{4\pi n}}{3\pi}+\int^{k_{F-}}_{0}ds \frac{s}{2\pi n}\left[\Sigma^F_s\left(\sqrt{s^2+4\pi n}\right)-\Sigma^F_s(s)\right],
\end{equation}
where the first contribution is the exchange energy of a circular Fermi sea and we have introduced: 
\begin{equation}
    \Sigma^{F}_s(k)=-\frac{2E_C}{\pi}\int^{\sqrt{s^2+4\pi n}}_{s}dk'  \frac{k'}{k+k'}K\left(\frac{4kk'}{(k+k')^2}\right).
\end{equation}
Next, we expand the self-energies for small $k_{F-}$: 
\begin{equation}\begin{split}
    \Sigma^F_s(s)&=-E_C\sqrt{4\pi n}+\frac{2E_C}{\pi}s-\frac{E_C}{4\sqrt{4\pi n}}s^2+O(s^4),\\
    \Sigma^F_s\left(\sqrt{s^2+4\pi n}\right)&=-\frac{2E_C\sqrt{4\pi n}}{\pi}+E_C\left(\frac{1}{2}-\frac{1}{\pi}\right)\frac{s^2}{\sqrt{4\pi n}}+O(s^4),
\end{split}\end{equation}
the resulting expression reads: 
\begin{equation}
    \Sigma^F_s\left(\sqrt{s^2+4\pi n}\right)-\Sigma^F_s(s)=E_C\sqrt{4\pi n}\left(1-\frac{2}{\pi}\right)-\frac{2E_C}{\pi}s+E_C\left(\frac{3}{4}-\frac{1}{\pi}\right)\frac{s^2}{\sqrt{4\pi n}}+O(s^4).
\end{equation}
Finally, we obtain: 
\begin{equation}
    \epsilon_{\rm int}[\hat n]\approx -\sqrt{4\pi n}\frac{4 E_C }{3\pi}+\frac{E_C}{\sqrt{4\pi n}}\left(1-\frac{2}{\pi}\right)k^2_{F-}-\frac{E_C}{3\pi^2 n}k^3_{F-}+\frac{E_C}{(4\pi n)^{3/2}}\left(\frac{3}{8}-\frac{1}{2\pi}\right)k^4_{F-}.
\end{equation}
In the regime $k_{F-}\ll\sqrt{4\pi n}$, The energy reads: 
\begin{equation}\begin{split}
    \epsilon=&-2\pi \alpha n+\frac{16\pi^2}{3}\beta n^2-\sqrt{4\pi n}\frac{4 E_C }{3\pi}
    +k_{F-}^2\left[\frac{E_C}{\sqrt{4\pi n}}\left(1-\frac{2}{\pi}\right)-\alpha+4\pi n\beta\right]\\
    &-k^3_{F-}\frac{E_C}{3\pi^2n}+k^4_{F-}\left[\beta+\frac{E_C}{(4\pi n)^{3/2}}\left(\frac{3}{8}-\frac{1}{2\pi}\right)\right].
\end{split}\end{equation}
Coulomb exchange offsets the negative quadratic term in the bare dispersion, yielding the coefficient
\begin{equation}
r(n)=\frac{E_C}{\sqrt{4\pi n}}\left(1-\frac{2}{\pi}\right)-\alpha+4\pi n\beta.
\end{equation}
This coefficient determines the local stability of the simply connected Fermi sea: opening an infinitesimal inner hole pocket raises the energy for $r(n)>0$ and lowers it for $r(n)<0$. Exchange, therefore, stabilizes the simply connected state, shifting the local-stability boundary $r(n)=0$ toward larger values of $\alpha/(4\pi\beta)$.

Finally, quantum geometry can further favor the suppression of the inner hole pocket. First, reduced overlaps between Bloch states weaken the exchange-energy gain in Eq.~\eqref{app:hf}, favoring occupation of momentum-space regions with larger mutual overlaps, associated locally with a smaller quantum metric.
Second, the quantum metric in R$L$G peaks near the outer rim of the Mexican-hat dispersion, making this exchange penalty particularly pronounced there and favoring a redistribution toward the center, which suppresses the inner hole pocket.

\end{document}